\PassOptionsToPackage{unicode}{hyperref}
\PassOptionsToPackage{hyphens}{url}
\PassOptionsToPackage{dvipsnames,svgnames,x11names}{xcolor}
\documentclass[
  12pt]{article}

\usepackage{amsmath,amssymb}
\usepackage{multirow} 
\usepackage{geometry} 
\usepackage{siunitx}
\usepackage[ruled,vlined]{algorithm2e}
\usepackage{iftex}
\ifPDFTeX
  \usepackage[T1]{fontenc}
  \usepackage[utf8]{inputenc}
  \usepackage{textcomp} 
\else 
  \usepackage{unicode-math}
  \defaultfontfeatures{Scale=MatchLowercase}
  \defaultfontfeatures[\rmfamily]{Ligatures=TeX,Scale=1}
\fi
\usepackage{lmodern}
\ifPDFTeX\else  
\fi
\IfFileExists{upquote.sty}{\usepackage{upquote}}{}
\IfFileExists{microtype.sty}{
  \usepackage[]{microtype}
  \UseMicrotypeSet[protrusion]{basicmath} 
}{}
\makeatletter
\@ifundefined{KOMAClassName}{
  \IfFileExists{parskip.sty}{%
    \usepackage{parskip}
  }{
    \setlength{\parindent}{0pt}
    \setlength{\parskip}{6pt plus 2pt minus 1pt}}
}{
  \KOMAoptions{parskip=half}}
\makeatother
\usepackage{xcolor}
\makeatletter
\ifx\paragraph\undefined\else
  \let\oldparagraph\paragraph
  \renewcommand{\paragraph}{
    \@ifstar
      \xxxParagraphStar
      \xxxParagraphNoStar
  }
  \newcommand{\xxxParagraphStar}[1]{\oldparagraph*{#1}\mbox{}}
  \newcommand{\xxxParagraphNoStar}[1]{\oldparagraph{#1}\mbox{}}
\fi
\ifx\subparagraph\undefined\else
  \let\oldsubparagraph\subparagraph
  \renewcommand{\subparagraph}{
    \@ifstar
      \xxxSubParagraphStar
      \xxxSubParagraphNoStar
  }
  \newcommand{\xxxSubParagraphStar}[1]{\oldsubparagraph*{#1}\mbox{}}
  \newcommand{\xxxSubParagraphNoStar}[1]{\oldsubparagraph{#1}\mbox{}}
\fi
\makeatother

\providecommand{\tightlist}{%
  \setlength{\itemsep}{0pt}\setlength{\parskip}{0pt}}\usepackage{longtable,booktabs,array}
\usepackage{calc} 
\usepackage{etoolbox}
\makeatletter
\patchcmd\longtable{\par}{\if@noskipsec\mbox{}\fi\par}{}{}
\makeatother
\IfFileExists{footnotehyper.sty}{\usepackage{footnotehyper}}{\usepackage{footnote}}
\makesavenoteenv{longtable}
\usepackage{graphicx}
\makeatletter
\def\maxwidth{\ifdim\Gin@nat@width>\linewidth\linewidth\else\Gin@nat@width\fi}
\def\maxheight{\ifdim\Gin@nat@height>\textheight\textheight\else\Gin@nat@height\fi}
\makeatother
\setkeys{Gin}{width=\maxwidth,height=\maxheight,keepaspectratio}
\makeatletter
\def\fps@figure{htbp}
\makeatother

\makeatletter
\@ifpackageloaded{caption}{}{\usepackage{caption}}
\AtBeginDocument{%
\ifdefined\contentsname
  \renewcommand*\contentsname{Table of contents}
\else
  \newcommand\contentsname{Table of contents}
\fi
\ifdefined\listfigurename
  \renewcommand*\listfigurename{List of Figures}
\else
  \newcommand\listfigurename{List of Figures}
\fi
\ifdefined\listtablename
  \renewcommand*\listtablename{List of Tables}
\else
  \newcommand\listtablename{List of Tables}
\fi
\ifdefined\figurename
  \renewcommand*\figurename{Figure}
\else
  \newcommand\figurename{Figure}
\fi
\ifdefined\tablename
  \renewcommand*\tablename{Table}
\else
  \newcommand\tablename{Table}
\fi
}
\@ifpackageloaded{float}{}{\usepackage{float}}
\floatstyle{ruled}
\@ifundefined{c@chapter}{\newfloat{codelisting}{h}{lop}}{\newfloat{codelisting}{h}{lop}[chapter]}
\floatname{codelisting}{Listing}

\makeatother
\makeatletter
\@ifpackageloaded{caption}{}{\usepackage{caption}}
\@ifpackageloaded{subcaption}{}{\usepackage{subcaption}}
\makeatother

\ifLuaTeX
  \usepackage{selnolig}  
\fi
\usepackage[]{natbib}
\usepackage{bookmark}

\IfFileExists{xurl.sty}{\usepackage{xurl}}{} 
\hypersetup{
  pdftitle={Title},
  pdfauthor={Author 1; Author 2},
  pdfkeywords={3 to 6 keywords, that do not appear in the title},
  colorlinks=true,
  linkcolor={blue},
  filecolor={Maroon},
  citecolor={Blue},
  urlcolor={Blue},
  pdfcreator={LaTeX via pandoc}}

\newcommand{\anon}{1}

\newtheorem{theorem}{Theorem}[section]

\newtheorem{lemma}[theorem]{Lemma}

\begin{document}

\def\spacingset#1{\renewcommand{\baselinestretch}%
{#1}\small\normalsize} \spacingset{1}


\if1\anon
{
  \title{\bf Bayesian factorization via $L_{1/2}$ shrinkage}
  \author{Shicheng Liu, Qingping Zhou, Yanan Fan, and Xiongwen Ke
 \thanks{
    Corresponding author: kexiongwen@csu.edu.cn}\hspace{.2cm}      \\
    School of Mathematics and Statistics, Central South University\\
    Data61, CSIRO, Sydney, Australia
   }
  \maketitle
} \fi

\if0\anon
{
  \bigskip
  \bigskip
  \bigskip
  \begin{center}
    {\LARGE\bf Bayesian factorization via $L_{1/2}$ shrinkage}
\end{center}
  \medskip
} \fi

\bigskip
\begin{abstract}
Factor models are widely used for dimension reduction. Bayesian approaches to these models often place a prior on the factor loadings that allows for infinitely many factors, with loadings increasingly shrunk toward zero as the column index increases. However, existing increasing shrinkage priors often possess complex hierarchical structures that complicate posterior inference. To address this issue, we propose using an $L_{1/2}$ shrinkage prior. We demonstrate that by carefully setting the parameters in the hyper prior of its global shrinkage parameters, the increasing shrinkage property is preserved. Our prior specification is simple, facilitating the construction of an efficient Gibbs sampler for exact posterior inference. For faster computation, we also propose a variational approximation algorithm. Through numerical studies, we compare our approaches with current popular Bayesian methods for factor models, demonstrating their merits in terms of accuracy and computational efficiency.
\end{abstract}

\noindent%
\vfill

\newpage
\spacingset{1.8}

\section{Introduction}\label{sec:intro}

The factor model can represent high-dimensional data through latent factors by performing a low-rank factorization of the covariance matrix. It has been applied in various fields, including social sciences \citep{samartsidis2020bayesian}, genomics studies \citep{carvalho2008high}, economics \citep{korobilis2025probabilistic}, and computer vision \citep{lee2005acquiring}. However, additional structural assumptions are often necessary to address the challenges associated with this model. For instance, the dimension of the latent factors is typically unknown, and it is natural to expect that the additional dimensions of the latent factor will play a progressively less important role in characterizing the structure of the data. Therefore, the associated factor should receive a stronger penalty. Furthermore, a series of works \citep{baik2006eigenvalues,johnstone2009consistency, jung2009pca,nadler2008finite,birnbaum2013minimax} have shown that the traditional principal component analysis (PCA) approach can lead to inconsistent estimation of the principal eigenvectors in the high-dimensional setting. The most popular approach to address this issue is to assume that the leading eigenvectors exhibit a certain type of sparsity, commonly referred to as sparse PCA in the literature \citep{zou2006sparse, d2008optimal, cai2013sparse}. 

On the Bayesian front, numerous priors have been developed in recent years to resolve these challenges in factor models. The main idea is to induce the group sparsity among the columns in the loading factor matrix, which leads to an automatic selection of the factor dimension and shrinking elements of the loading matrix toward zero. The seminal work by \cite{bhattacharya2011sparse} introduced the multiplicative gamma process (MGP) shrinkage prior, which allows for infinitely many factors in the model, with elements of the factor loading matrix increasingly shrunk toward zero as the column index increases. Consequently, posterior sampling adopts an adaptive approach to automatically select the latent dimension, facilitating the identification of the optimal dimensionality. Building on this idea, the spike-and-slab type prior, which assigns sufficient probability mass to zero, has been widely used to infer the factor dimensionality. Researchers have combined the spike-and-slab prior with the Poisson process \citep{pillai2014posterior},  the Indian buffet process \citep{knowles2011nonparametric,rovckova2016fast,ohn2022posterior}, and the stick-breaking process \citep{legramanti2020bayesian} to construct the prior with increasing shrinkage properties. Another spike-and-slab prior provided by \cite{gao2015rate} offers promising theoretical results but is computationally intractable; they successfully implemented a posterior sampler only for factor models with a one-dimensional factor.

Bayesian inference on the posterior often relies on Markov-Chain Monte Carlo (MCMC) algorithms. Recently, alternatives to MCMC as fast inference approaches have been proposed, including the expectation maximization algorithms for posterior mode search with the spike-and-slab Lasso-Indian buffet process prior (SSL-IBP) \citep{rovckova2016fast} and multi-scale generalized double Pareto prior \citep{srivastava2017expandable}, as well as the variational inference (VI) for posterior approximation with multiplicative gamma process shrinkage prior \citep{hansen2025fast}, and the spike-and-slab prior \citep{ning2024spike}. We see that most of the recent development of the prior for the Bayesian factor model used spike-and-slab type priors as a basic building block. Few follow-up works based on the global local shrinkage prior \citep{polson2010shrink} have emerged since the introduction of the MGP shrinkage prior by \cite{bhattacharya2011sparse}. \cite{srivastava2017expandable} proposed using the generalized double Pareto for factor models but focused solely on posterior mode search. 
 
 In this paper, we consider using $L_{1/2}$ shrinkage prior from \cite{ke2025bayesian} for the Bayesian factor model. By carefully setting the parameters in the prior of global shrinkage parameters, we ensure that the elements of the factor loading matrix are increasingly shrunk toward zero as the column index increases, achieving a stronger sense of shrinkage than the MGP prior. Additionally, this prior has a simple structure, easing the posterior inference. We construct a Gibbs sampler as an exact inference approach. Since MCMC methods are generally computationally intensive and memory-demanding, we also propose a novel collapsed variational inference method that integrates out both global and local shrinkage parameters for fast posterior approximation. In our numerical studies, we demonstrate the performance of our approach on both simulated and real datasets.
 
 The rest of the paper is organized as follows. Section 2 introduces the prior setting and discusses its properties. Section 3 details the Gibbs sampler. Section 4 presents the variational inference algorithm. Section 5 evaluates the proposed algorithms through numerical studies. Section 6 concludes with discussion.

\section{Model and priors}\label{sec:prior}

This section introduces the latent factor model, followed by the specification of the $L_{1/2}$ shrinkage prior and its properties.

\subsection{Latent factor model}

The setup for factor analysis consists of an $n \times p$ matrix $\boldsymbol{X} = (\boldsymbol{x}_{1},..., \boldsymbol{x}_{n})^{\top}$ of $n$ independent observations with dimension $p$. The latent factor model is of the form
\begin{equation}\label{eq:LFM}
\boldsymbol{x}_{i}=\boldsymbol{B} \boldsymbol{\eta}_{i}+ \boldsymbol{\epsilon}_i, \quad  \boldsymbol{\eta}_i \stackrel{\text { i.i.d. }}{\sim} \mathcal{N}_{r}\left(\boldsymbol{0}, \boldsymbol{I}_r\right), \quad \epsilon_{i}  \stackrel{\text { i.i.d. }}{\sim}  \mathcal{N}_{p}(\boldsymbol{0},\boldsymbol{\Omega}) \quad (i=1,...,n),
\end{equation}
where $\boldsymbol{B}$ is the $p \times K$ factor loading matrix that weights the contributions of the $K$- dimensional latent factor $\boldsymbol{\eta}_{i}$ and $\boldsymbol{\epsilon}_{i}$ is an idiosyncratic error with covariance $\boldsymbol{\Omega} = \operatorname{diag}(\sigma_{1}^{2},...,\sigma_{p}^{2})$. Marginalizing out the latent factor $\boldsymbol{\eta}_{i}$, we have $f(\boldsymbol{x}_{i} \mid \boldsymbol{B},\boldsymbol{\Omega})=\mathcal{N}(\boldsymbol{0},\boldsymbol{B}\boldsymbol{B}^{\top}+\boldsymbol{\Omega})$. 

\subsection{Prior Specification}
Motivated by the Bayesian infinite factor model \citep{bhattacharya2011sparse,legramanti2020bayesian}, our approach begins by assigning $L_{1/2}$ shrinkage prior \citep{ke2025bayesian} on the individual elements in $\boldsymbol{B} = \left\{B_{jk}\right\}$ with $j=1,...,p$ and $k=1,2,...,+\infty$ and making the global shrinkage parameter stochastically increase with the column index of the factor loading matrix. Since the dimension of the latent factor $K$ is usually unknown, this setting allows the posterior parameter space $\boldsymbol{B}$ to contain all possible numbers of latent factors. More formally,
\begin{equation}\label{eq:prior}
\pi(B_{jk} \mid \lambda_{k}) =\frac{\lambda_{k}^{2}}{4}\exp(-\lambda_{k}|B_{jk}|^{\frac{1}{2}}), \quad  \lambda_{k} \sim \operatorname{Gamma}\left(a+k^{c_{1}}, \frac{1}{k^{c_{2}}}\right), 
\end{equation}
where $\lambda_{k}$ is a global shrinkage parameter for the $k$th column and $a>0$, $c_{1}>0$, $c_{2}>0$. For the $L_{1/2}$ shrinkage prior, it has the Gaussian mixture representation as shown by \cite{ke2025bayesian}:
\begin{equation}\label{eq:GM}
\begin{aligned}
& B_{jk} \mid \tau_{jk}^2, \lambda_{k} \sim N\left(0, \frac{\tau_{jk}^2}{\lambda_{k}^{4}}\right), \quad \tau_{jk}^2 \mid v_{jk} \sim \operatorname{Exp}\left(\frac{1}{2 v_{jk}^2}\right), \\
& v_{jk} \sim \operatorname{Gamma}\left(\frac{3}{2}, \frac{1}{4}\right).
\end{aligned}
\end{equation}
Finally, an inverse-gamma prior is placed on the diagonal entries of $\boldsymbol{\Omega}$: $\sigma_{j}^{2} \sim \operatorname{IG}(a_{\sigma},b_{\sigma})$ for $j = 1,...,p$.

\subsection{Properties of the shrinkage prior}

We start with a prior for a loading matrix $\boldsymbol{B} \in \mathbb{R}^{p \times \infty}$ having infinitely many columns. It was shown by \cite{bhattacharya2011sparse} that the set of loading matrix that leads to well-defined covariance matrices is
$$
\Theta_{\boldsymbol{B}}=\left\{\boldsymbol{B} =\left(B_{j k}\right), \,\, j=1, \ldots, p,\,\, k=1 \ldots, +\infty,\,\, \max _{1 \leq j \leq p} \sum_{h=1}^{\infty} B_{j k}^2<\infty\right\}.
$$
The factor loading matrix $\boldsymbol{B}$ in the set $\Theta_{\boldsymbol{B}}$ guarantees that $\boldsymbol{\Sigma} = \boldsymbol{B}\boldsymbol{B}^{\top}+\boldsymbol{\Omega}$ has finite entries and is positive semi-definite. The next lemma provides the condition that the prior measure $\mathbb{P}$ has full support on $\Theta_{\boldsymbol{B}}$.
\begin{lemma}\label{lem:Well_defined_covariance}
If $a \geq4$ and $c_{1} + c_{2} > \frac{1}{4}$, then $\mathbb{P}(\Theta_{\boldsymbol{B}}) = 1$. 
\end{lemma}

The proof is given in the Appendix A supplementary, along with the other proofs in this section. In practice, the loading matrix $\boldsymbol{B}$ is truncated to a finite number of columns for tractable computation, as in \cite{bhattacharya2011sparse}. The truncated loading matrix $\boldsymbol{B_{K}}$ is obtained by deleting the columns of $\boldsymbol{B}$ with index $k>K$. The next theorem guarantees that the choice of $K$ can make the covariance matrix $\boldsymbol{\Sigma}_{K} = \boldsymbol{B_{K}}\boldsymbol{B_{K}}^{\top}+\boldsymbol{\Omega}$ arbitrarily close to $\boldsymbol{\Sigma} = \boldsymbol{B}\boldsymbol{B}^{\top}+\boldsymbol{\Omega}$, where distance between $\boldsymbol{B_{K}}$ and $\boldsymbol{B}$ is measured using the sup-norm.

\begin{theorem}\label{thm:truncation}
If $a \geq 4$ and $c_{1} + c_{2} > \frac{1}{4}$, then for any $\epsilon>0$,
\begin{equation}\label{eq:inverse_power_rate}
\mathbb{P}(d_{\infty}(\boldsymbol{\Sigma},\boldsymbol{\Sigma_{K}}) > \epsilon) \leq \frac{2Hp}{\epsilon (K+1)^{4(c_{1}+c_{2})}}
\end{equation}
for $K > (\frac{4H}{3\epsilon})^{\frac{1}{4(c_{1}+c_{2})}}-1$, where $H =  \frac{480(c_{1}+c_{2})}{4(c_{1}+c_{2})-1}$.
\end{theorem}

The above theorem states that the prior probability of $\boldsymbol{\Sigma}_{K}$ being arbitrarily close to $\boldsymbol{\Sigma}$ with inverse power fast rate as $K \rightarrow +\infty$. \cite{durante2017note} showed that the increasing shrinkage properties of the MGP shrinkage prior only hold in expectation but not in probability. With this motivation, the cumulative shrinkage process (CSP) prior \citep{legramanti2020bayesian} was proposed. The next lemma states that this stronger sense of increasing shrinkage property also holds under the prior setting (\ref{eq:prior}).

\begin{lemma}\label{lem:stochastic_increasing_shrinkage}
$\mathbb{P}(|B_{j,k+1}| \leq \epsilon) >  \mathbb{P}(|B_{j,k}| \leq \epsilon)$ for any $\epsilon>0$ and $k \in \mathbb{N}^{+}$.
\end{lemma}

Finally, we show that our prior assigns positive probability to arbitrarily small neighborhoods around any covariance matrix.

\begin{theorem}\label{thm:prior_support}
Let $\mathcal{B}_{\epsilon}^{\infty}\left(\boldsymbol{\Sigma}_{0}\right) = \left\{\boldsymbol{\Sigma} \in \boldsymbol{S}^{p}: \boldsymbol{x}^{\top}\boldsymbol{\Sigma}\boldsymbol{x} \geq 0 \,\, \text{for all} \,\, x\neq 0 \,\, \text{and} \,\, d_{\infty}(\boldsymbol{\Sigma},\boldsymbol{\Sigma}_{0})\leq \epsilon     \right\}$	be an $\epsilon$-neighborhood of $\boldsymbol{\Sigma}_{0}$ with radius of $\epsilon$ under the sup-norm. If $\boldsymbol{\Sigma}_{0}$ is any $p \times p$  covariance matrix, then if $a \geq 4$ and $c_{1}+c_{2}>\frac{1}{4}$, $\mathbb{P}\left\{\mathcal{B}_{\epsilon}^{\infty}\left(\boldsymbol{\Sigma}_{0}\right)\right\}>0$ for any $\epsilon>0$.
\end{theorem}
Then by using Theorem 2 of \cite{bhattacharya2011sparse}, we obtain the weak consistency of the posterior for $\boldsymbol{\Sigma}$.

\section{Gibbs Sampling}\label{sec:Gibbs}

In this section, we denote $\boldsymbol{B}_{j \cdot}$ as the $1 \times K$ row vector and $\boldsymbol{B}_{\cdot k}$ as the $p \times 1$ column vector from the matrix $\boldsymbol{B}$. We further denote $\boldsymbol{X}_{\cdot j}$ as the $n \times 1$ column vector from the matrix $\boldsymbol{X}$. Such notations will be used in the remaining sections. Then we propose a Gibbs sampler for the posterior computation:
\begin{enumerate}
\tightlist
\item Sample $\sigma_{j}^{2}$ for $j = 1,...,p$ independently from 
$$
\sigma_{j}^{2} \mid \boldsymbol{x}_{j \cdot},\boldsymbol{B}_{j \cdot}, \boldsymbol{\eta} \sim \operatorname{InvGamma}\left(a_{\sigma}+\frac{n}{2},b_{\sigma}+\frac{1}{2}\|\boldsymbol{X}_{\cdot j}^{\top}-\boldsymbol{B}_{j \cdot}\boldsymbol{\eta}\|^{2}\right).
$$
\item Sample $\boldsymbol{B}_{j \cdot}$ for $j=1,...,p$ independently from
$$
\boldsymbol{B}_{j \cdot} \mid \boldsymbol{x}_{j \cdot},\boldsymbol{\eta}, \boldsymbol{\tau}_{j }^{2},\sigma_{j}^{2},\boldsymbol{\lambda} \sim \mathcal{N}_{r}\left(\left(\frac{\boldsymbol{\eta} \boldsymbol{\eta}^{\top}}{\sigma_{j}^{2}}+\operatorname{diag}\left(\frac{\boldsymbol{\lambda}^{4}}{\boldsymbol{\tau}_{j}^{2}}\right)\right)^{-1}\frac{\boldsymbol{\eta}\boldsymbol{X}_{\cdot j}}{\sigma_{j}^{2}},\left(\frac{\boldsymbol{\eta} \boldsymbol{\eta}^{\top}}{\sigma_{j}^{2}}+\operatorname{diag}\left(\frac{\boldsymbol{\lambda}^{4}}{\boldsymbol{\tau}_{j}^{2}}\right)\right)^{-1}\right).
$$
\item Sample $(\boldsymbol{\eta},\boldsymbol{\tau}^{2},\boldsymbol{\lambda})$ from
$$\pi(\boldsymbol{\eta},\boldsymbol{\tau}^{2},\boldsymbol{\lambda} \mid \boldsymbol{B},\boldsymbol{X},\boldsymbol{\Omega}) = \left[\prod_{j,k}\pi(\tau_{jk}^{2} \mid \lambda_{k}, B_{jk})\prod_{k=1}^{K}\pi(\lambda_{k} \mid \boldsymbol{B}_{\cdot k})\right] \prod_{i=1}^{n}\pi(\boldsymbol{\eta}_{i} \mid \boldsymbol{B}, \boldsymbol{\Omega},\boldsymbol{x}_{i}),$$
which is done by the following step: 
\begin{itemize}
\tightlist
    \item \textbf{Sample} $\boldsymbol{\eta}_{i}$ for $i = 1,...,n$ independently from
$$
\boldsymbol{\eta}_{i} \mid \boldsymbol{B}, \boldsymbol{\Omega},\boldsymbol{x}_{i} \sim    \mathcal{N}_{K}\left((\boldsymbol{I}_{K}+\boldsymbol{B}^{\top}\boldsymbol{\Omega}^{-1} \boldsymbol{B})^{-1}\boldsymbol{B}^{\top}\boldsymbol{\Omega}^{-1} \boldsymbol{x}_{i}, (\boldsymbol{I}_{K}+\boldsymbol{B}^{\top}\boldsymbol{\Omega}^{-1} \boldsymbol{B})^{-1}\right)
$$ 
\item \textbf{Sample} $(\boldsymbol{\lambda},\boldsymbol{\tau}^{2})$ sequentially from 
\begin{itemize}
\tightlist
       \item  \textbf{Sample} $\boldsymbol{\lambda}$ for $k = 1,...,K$ independently from
       $$\lambda_{k} \mid \boldsymbol{B}_{k} \sim\operatorname{Gamma}\left(2 p  + a + k^{c_{1}}, \sum_{j=1}^{p}|B_{jk}|^{\frac{1}{2}}+\frac{1}{k^{c_{2}}}\right).$$ 
       \item \textbf{Sample} $\tau_{jk}^{2}$ for $j = 1,...,p$ and $k = 1,...,K$ independently from
       
       $\pi(\tau_{jk}^{2} \mid \lambda_{k}, B_{jk})$, which is done by sampling $\pi(\tau_{jk}^{2}, v_{jk} \mid \lambda_{k}, B_{jk})$:
       \begin{itemize}
       \item \text{Sample} $\frac{1}{v_{jk}} \mid B_{jk}, \lambda_{k} \sim \operatorname{InvGaussian}\left(\frac{1}{2\lambda_{k}|B_{jk}|^{\frac{1}{2}}},\frac{1}{2}\right)$
       \item \text{Sample} $\frac{1}{\tau_{jk}^{2}} \mid v_{jk},B_{jk},\lambda_{k} \sim \operatorname{InvGaussian}\left(\frac{1}{\lambda_{k}^{2}v_{jk}|B|_{jk}},\frac{1}{v_{jk}^{2}}\right)$
       \end{itemize}
\end{itemize}
\end{itemize}

\end{enumerate}
We observe that for different latent dimensions $k$, the conditional posteriors of global and local shrinkage parameters are independent. In contrast, this property does not hold for the posteriors derived from the MGP and CSP priors. 

\section{Variational inference}\label{sec:VI}
Although the Gibbs sampler can provide an exact posterior inference, it is computationally intensive and memory-demanding. In this section, we use the variational inference to approximate the joint posterior $\pi(\boldsymbol{B},\boldsymbol{\Omega}, \boldsymbol{\eta} \mid \boldsymbol{X})$.

\subsection{Collapsed variational inference}
We consider the following mean-field assumption:
\begin{equation}\label{eq:collapsed_mean_field}
q(\boldsymbol{\boldsymbol{B}},\boldsymbol{\Omega},\boldsymbol{\eta})	 = q(\boldsymbol{\Omega}) q(\boldsymbol{\eta})q(\boldsymbol{B}) = q(\boldsymbol{B})\prod_{j=1}^{p}q_{j}(\sigma_{j}^{2}) \prod_{i=1}^{n} q_{i}(\boldsymbol{\eta}_{i}),
\end{equation}
where the second equality is followed by the fact that if the conditional posterior of $\boldsymbol{\Omega}$ and $\boldsymbol{\eta}$  have an independent structure, then the independent structure also holds for their variational posterior \citep{bishop2006pattern}.  Our approach relates to the collapsed variational inference \citep{teh2007collapsed}, which aims to improve the accuracy of the approximation by marginalizing out the global and local shrinkage parameters. The evidence lower bound (ELBO) is then:
\begin{equation}\label{eq:ELBO}
\begin{aligned}
& \mathcal{L}(q(\boldsymbol{B}), q(\boldsymbol{\eta}),q(\boldsymbol{\sigma}^{2})) \\
 = & \mathbb{E}_{q(\boldsymbol{\Omega})q(\boldsymbol{B})q(\boldsymbol{\eta})}[\log p(\boldsymbol{X} \mid \boldsymbol{B},\boldsymbol{\Omega},\boldsymbol{\eta})]+
  \mathbb{E}_{q(\boldsymbol{\eta})}[\log \pi(\boldsymbol{\eta})] +\mathbb{E}_{q(\boldsymbol{B})}[\log \pi( \boldsymbol{B})] +\mathbb{E}_{q(\boldsymbol{\Omega})}[\log \pi(\boldsymbol{\Omega})] \\
&    - \mathbb{E}_{q(\boldsymbol{\eta})}[\log q(\boldsymbol{\eta})]  - \mathbb{E}_{q(\boldsymbol{B})}[\log q(\boldsymbol{B})] - \mathbb{E}_{q(\boldsymbol{\Omega})}[\log q(\boldsymbol{\Omega})]
\end{aligned}
\end{equation}
From section \ref{sec:Gibbs}, we see that both $\pi(\boldsymbol{\eta} \mid \boldsymbol{\Omega},\boldsymbol{X},\boldsymbol{B})$, and $\pi(\boldsymbol{\Omega} \mid \boldsymbol{X},\boldsymbol{B},\boldsymbol{\eta})$ belong to the exponential family. Therefore, maximizing the ELBO of (\ref{eq:ELBO}) leads to the  $q(\boldsymbol{\eta})$ and $q(\boldsymbol{\sigma}^{2})$ belong to the same distribution as their conditional posterior \citep{bishop2006pattern}. Thus, we have:
\begin{equation}
\begin{aligned}
q_{j}(\sigma_{j}^{2}) & \sim \operatorname{InvGamma}\left(\alpha_{\sigma}+\frac{n}{2}, b_{\sigma}+\frac{1}{2}\|\boldsymbol{X}_{\cdot j}^{\top} - \boldsymbol{\mu}_{j}\boldsymbol{w} \|^{2} +\frac{n}{2}\boldsymbol{\mu}_{j}\Phi \boldsymbol{\mu}_{j}^{\top} +\frac{1}{2}\operatorname{Trc}\left[\operatorname{Cov}(\boldsymbol{B}_{j \cdot})(\boldsymbol{w}\boldsymbol{w}^{\top}+n\Phi)\right] \right)\\
q_{i}(\boldsymbol{\eta}_{i}) & \sim \mathcal{N}_{K}( (\boldsymbol{I}_{K} + \boldsymbol{\mu}^{\top}\boldsymbol{C}\boldsymbol{\mu}+ \boldsymbol{A})^{-1}\boldsymbol{\mu}^{\top}\boldsymbol{C}\boldsymbol{x}_{i},(\boldsymbol{I}_{K} + \boldsymbol{\mu}^{\top}\boldsymbol{C}\boldsymbol{\mu}+ \boldsymbol{A})^{-1}),
\end{aligned}
\end{equation}
where $\boldsymbol{\mu} = \mathbb{E}_{q(\boldsymbol{B})}[\boldsymbol{B}]$, $\boldsymbol{w}=\mathbb{E}_{q(\boldsymbol{\eta})}[\boldsymbol{\eta}]$, $\boldsymbol{C}=\operatorname{diag}\left(\left\{\mathbb{E}_{q_{j}(\sigma_{j}^{2})}\left[\frac{1}{\sigma_{j}^{2}}\right]\right\}_{j=1}^{p}\right)$, $\boldsymbol{A} =  \sum_{j=1}^{p} \mathbb{E}_{q_{j}(\sigma_{j}^{2})}\left[\frac{1}{\sigma_{j}^{2}}\right]\operatorname{Cov}(\boldsymbol{B}_{j \cdot})$ and $\Phi = \operatorname{Cov}_{q_{i}(\boldsymbol{\eta}_{i})}[\boldsymbol{\eta}_{i}]= (\boldsymbol{I}_{K} + \boldsymbol{\mu}^{\top}\boldsymbol{C}\boldsymbol{\mu}+ \boldsymbol{A})^{-1}$.

To maximize the ELBO in (\ref{eq:ELBO}), we perform coordinate-ascent optimization, iteratively updating the variational parameters in each variational factor while keeping others fixed.

\subsection{Optimizing the ELBO with respect to \texorpdfstring{$q(\boldsymbol{B})$}{q(B)}}
Since the conditional posterior $\pi(\boldsymbol{B} \mid \boldsymbol{\eta}, \boldsymbol{X}, \boldsymbol{\Omega})$ does not belong to any simple distribution family, neither does $q(\boldsymbol{B})$. Thus additional effort is required to make computation tractable. We further assume that $q(\boldsymbol{B}) = \prod_{j=1}^{p}q_{j}(\boldsymbol{B}_{j \cdot})$ and consider the variational posterior
$$
\mathcal{Q}^{\mathrm{MF}} = \left\{q_{j}(\boldsymbol{B}_{j \cdot}):=t_\nu(\boldsymbol{\mu}_{j}, \boldsymbol{\Sigma}_{j})\right\},
$$
where $t_\nu(\boldsymbol{\mu}_{j}, \boldsymbol{\Sigma}_{j})$ is the $K$-dimensional multivariate t-distribution with location parameter $\boldsymbol{\mu}_{j}$, scale matrix $\boldsymbol{\Sigma}_{j}$ and $\nu>0$ degrees of freedom. When $\nu \rightarrow +\infty$, the multivariate t-distribution converges to the Gaussian distribution. We set $\boldsymbol{\varphi}_{j}=(\boldsymbol{\mu}_{j},\boldsymbol{\Lambda}_{j},\nu_{j})$ as variational parameters with $\boldsymbol{\Lambda}_{j} = \boldsymbol{\Sigma}_{j}^{-1}$ for optimization. We further restrict $\nu_{j}>2$ to guarantee the existence of the mean and covariance of the multivariate t-distribution. Before deriving our algorithm,  we state some useful results:
\begin{itemize}
\tightlist
    \item The multivariate t-distribution has a scale-mixture of Gaussian hierarchical representation: 
    \begin{equation}\label{eq:Gaussian_Gamma}
    q_{j}(\boldsymbol{B}_{j \cdot} \mid u_{j}) \sim \mathcal{N}_{K}\left(\boldsymbol{\mu}_{j}, u_{j}^{-1} \boldsymbol{\Lambda}_{j}^{-1}\right), \quad q_{j}(u_{j}) \sim \operatorname{Gamma}(\nu_{j} / 2, \nu_{j} / 2).
    \end{equation}
    \item The entropy of $K$-dimensional multivariate t-distribution is 
    \begin{equation}\label{eq:entropy_mul_t}
    -\mathbb{E}_{q_{j}(\boldsymbol{B}_{j \cdot})}[\log q(\boldsymbol{B}_{j \cdot})]=-\frac{1}{2} \ln |\boldsymbol{\Lambda}_{j}| +h(\nu_{j}),
    \end{equation}
    where $h(\nu_{j})= \log \Gamma\left(\frac{\nu_{j}}{2}\right)-\log \Gamma\left(\frac{\nu_{j}+K}{2}\right)+\frac{K}{2} \log (\nu_{j} \pi)+\frac{\nu+K}{2}\left[\psi\left(\frac{\nu_{j}+K}{2}\right)-\psi\left(\frac{\nu_{j}}{2}\right)\right]$ and $\psi(\cdot)$ is the digamma function.
    \item Simple extension of Bonnet’s and Price’s Gradients \citep{opper2009variational}: Consider the gradients with respect to the $\boldsymbol{\mu}$ and $\boldsymbol{\Sigma}$ of a Gaussian distribution $\mathcal{N}\left(\boldsymbol{\mu}, u^{-1}\boldsymbol{\Sigma}\right)$, then for some suitable smooth function $f(\cdot)$, we have
    \begin{equation}\label{eq:Bonnet}
    \begin{aligned} \nabla_{\boldsymbol{\mu}} \mathbb{E}_{\mathcal{N}\left(\boldsymbol{\mu}, u^{-1}\boldsymbol{\Sigma}\right)}[f(\mathbf{z})] & =\mathbb{E}_{\mathcal{N}\left(\boldsymbol{\mu}, u^{-1}\boldsymbol{\Sigma}\right)}\left[\nabla_{\mathbf{z}} f( \mathbf{z})\right], \\ \nabla_{\boldsymbol{\Sigma}} \mathbb{E}_{\mathcal{N}\left(\boldsymbol{\mu}, u^{-1}\boldsymbol{\Sigma}\right)}[f(\mathbf{z})] & =\frac{1}{2u} \mathbb{E}_{\mathcal{N}\left(\boldsymbol{\mu}, u^{-1}\boldsymbol{\Sigma}\right)}\left[\nabla_{\mathbf{z}}^2 f (\mathbf{z})\right] .
    \end{aligned}
    \end{equation}
\end{itemize}
Our strategy is to learn $\left\{\mu_{j}, \Lambda_j\right\}_{j=1}^p$ and $\left\{\nu_j\right\}_{j=1}^p$ separately, rather than jointly learning the variational parameters $\left\{\phi_j=\right. \left.\left(\boldsymbol{\mu}_j, \boldsymbol{\Lambda}_j, \nu_j\right)\right\}_{j=1}^p$. That is when updating one of them, the other should be fixed.

\subsubsection{Learning \texorpdfstring{$\boldsymbol{\mu}_{j}$}{mu\_j} and \texorpdfstring{$\boldsymbol{\Lambda}_{j}$}{Lambda\_j}}

Now we write down the ELBO in (\ref{eq:ELBO}) that depends only on $\boldsymbol{\mu}_{j}$ and $\boldsymbol{\Lambda}_{j}$:
\begin{equation}\label{eq:ELBO_B}
\mathcal{L}_{(\boldsymbol{\mu}_{j} ,\boldsymbol{\Lambda}_{j})}  =   - \frac{c_{j}}{2}\mathbb{E}_{q_{j}(\boldsymbol{B}_{j\cdot})}\left[\|\boldsymbol{X}_{\cdot j}^{\top} - \boldsymbol{B}_{j\cdot}\boldsymbol{w}\|^{2}+n\boldsymbol{B}_{j\cdot} \Phi\boldsymbol{B}_{j\cdot}^{\top}\right] + \mathbb{E}_{q(\boldsymbol{B} \mid \boldsymbol{u})q(\boldsymbol{u})}[\log \pi(\boldsymbol{B})] -\frac{1}{2} \ln |\boldsymbol{\Lambda}_{j}|,
\end{equation}
where $c_{j} = \mathbb{E}_{q_{j}(\sigma_{j}^{2})}\left[\frac{1}{\sigma_{j}^{2}}\right]$, $\boldsymbol{w}=\mathbb{E}_{q(\boldsymbol{\eta})}[\boldsymbol{\eta}]$ and $\Phi = \operatorname{Cov}_{q_{i}(\boldsymbol{\eta}_{i})}[\boldsymbol{\eta}_{i}]= (\boldsymbol{I}_{K} + \boldsymbol{\mu}^{\top}\boldsymbol{C}\boldsymbol{\mu}+ \boldsymbol{A})^{-1}$ with $\boldsymbol{\mu} = (\boldsymbol{\mu}_{1},...,\boldsymbol{\mu}_{p})^{\top}$.
It seems that we can obtain the Monte Carlo gradient of the ELBO with respect to the variational parameters by using Bonnet’s and Price’s theorem in equation (\ref{eq:Bonnet}). Then, the gradient descent type algorithm can be used. However, since $\log \pi(\boldsymbol{B})$ is non-differentiable and non-Lipschitz at zero \citep{ke2025bayesian}, this theorem can not be applied directly. The next Lemma helps us to find a tractable lower bound of (\ref{eq:ELBO_B}) as a surrogate function for optimization at each iteration.
\begin{lemma}\label{lem:lower_bound}
For any $\boldsymbol{B}^{*} \in \mathbb{R}^{p \times K}$, we have
 \begin{equation}\label{eq:MM_bound}
 \begin{aligned}
 \mathbb{E}_{q(\boldsymbol{B})}[\log \pi(\boldsymbol{B})] & \geq \mathbb{E}_{q(\boldsymbol{B})}\left\{\mathbb{E}_{\pi(\boldsymbol{\tau}^{2},\lambda_{k} \mid \boldsymbol{B}^{*})}[\log \pi(\boldsymbol{B} \mid \boldsymbol{\tau}^{2},\boldsymbol{\lambda})]\right\} + C\\
 & = -\frac{1}{2} \sum_{k,j}\mathbb{E}_{\pi(\tau_{jk}^{2},\lambda_{k} \mid B_{jk}^{*})}\left[\frac{\lambda_{k}^{4}}{\tau_{jk}^{2}}\right]\mathbb{E}_{q(B_{jk})}[B_{jk}^{2}] + C,
 \end{aligned}
 \end{equation}
where $C$ is the constant term that does not depend on the variational parameters $\boldsymbol{\mu}_{j}$ and $\boldsymbol{\Lambda}_{j}$. In addition, if $B_{jk}^{*} \neq 0$, 
\begin{equation}\label{eq:LQP}
\mathbb{E}_{\pi(\tau_{jk}^{2},\lambda_{k} \mid B_{jk}^{*})}\left[\frac{\lambda_{k}^{4}}{\tau_{jk}^{2}}\right] = \frac{(p+a/2+k^{c_{1}}/2)}{|B_{jk}^{*}|^{3/2}(\sum_{j=1}^{p}|B_{jk}^{*}|^{\frac{1}{2}}+b/k^{c_{2}})}.
\end{equation}
If $B_{jk}^{*} = 0$, then the lower bound in (\ref{eq:MM_bound}) becomes trivial as its right hand side attains $-\infty$.
\end{lemma}
The proof is given in the supplementary Appendix B, along with the other proofs in this section. By Lemma \ref{lem:lower_bound}, we have
\begin{equation}\label{eq:surrogate}
\begin{aligned}
\mathcal{L}(q_{j}(\boldsymbol{B}_{j\cdot}))  \geq \tilde{\mathcal{L}}(q_{j}(\boldsymbol{B}_{j\cdot})) & = - \frac{c_{j}}{2}\mathbb{E}_{q_{j}(\boldsymbol{B}_{j\cdot})}\left[\|\boldsymbol{X}_{\cdot j}^{\top} - \boldsymbol{B}_{j\cdot}\boldsymbol{w}\|^{2}+n\boldsymbol{B}_{j\cdot} \Phi\boldsymbol{B}_{j\cdot}^{\top}\right]\\
& -\frac{1}{2}\sum_{k=1}^{K}\mathbb{E}_{\pi(\tau_{jk}^{2},\lambda_{k} \mid B_{jk}^{*})}\left[\frac{\lambda_{k}^{4}}{\tau_{jk}^{2}}\right]\mathbb{E}_{q(B_{jk})}[B_{jk}^{2}]-\frac{1}{2} \ln |\boldsymbol{\Lambda}_{j}| + C.
\end{aligned}
\end{equation}
The lower bound we obtain in Lemma \ref{lem:lower_bound} is closely related to the local quadratic approximation \citep{fan2001variable}, which is one of the minorize-maximize (MM) algorithms \citep{hunter2005variable}. To avoid the trivial lower bound, \cite{hunter2005variable} suggested bounding the denominator away from zero by adding a small positive tolerance number to make a slightly perturbed version of the lower bound. In equation (\ref{eq:LQP}), we localize the lower bound (\ref{eq:surrogate}) by setting $B_{jk}^{*} = \mu_{jk,t-1}$ and modify the term $|\mu_{jk,t-1}|^{3/2}$ in the denominator to $\max(|\mu_{jk,t-1}|^{3/2}, 10^{-9})$ at iteration $t$. In practice, we found that this strategy works very well. Now, the gradient of variational parameters with respect to the surrogate function $\tilde{\mathcal{L}}(q_{j}(\boldsymbol{B}_{j\cdot}))$ can be obtained by Bonnet’s and Price’s theorem shown in (\ref{eq:Bonnet}):
\begin{equation}\label{eq:mc_gradient}
\begin{aligned}
\nabla_{\boldsymbol{\mu}_{j}}\tilde{\mathcal{L}}(q_{j}(\boldsymbol{B}_{j\cdot}))  & = c_{j}\boldsymbol{X}_{\cdot j}^{\top}-\boldsymbol{\mu}_{j} \left[c_{j}(\boldsymbol{w}\boldsymbol{w}^{\top}+ n\Phi)+\mathbb{E}_{\pi(\boldsymbol{\tau}_{j}^{2}, \boldsymbol{\lambda} \mid  \boldsymbol{B}_{j}^{*})}\left[\boldsymbol{\Lambda}_{0j}\right]\right],\\
\nabla_{\boldsymbol{\Lambda}_{j}}\tilde{\mathcal{L}}(q_{j}(\boldsymbol{B}_{j\cdot})) & = \frac{1}{2}\frac{\nu_{j}}{\nu_{j} -2}\boldsymbol{\Lambda}_{j}^{-1}\left[c_{j}(\boldsymbol{w}\boldsymbol{w}^{\top}+n \Phi)+\mathbb{E}_{\pi(\boldsymbol{\tau}_{j}^{2},\boldsymbol{\lambda} \mid  \boldsymbol{B}_{j}^{*})}\left[\boldsymbol{\Lambda}_{0j} \right]\right]\boldsymbol{\Lambda}_{j}^{-1} - \frac{1}{2}\boldsymbol{\Lambda}_{j}^{-1},
\end{aligned}
\end{equation}
where $c_{j} = \mathbb{E}_{q_{j}(\sigma_{j}^{2})}\left[\frac{1}{\sigma_{j}^{2}}\right]$, $\boldsymbol{w}=\mathbb{E}_{q(\boldsymbol{\eta})}[\boldsymbol{\eta}]$, $\Phi = \operatorname{Cov}_{q_{i}(\boldsymbol{\eta}_{i})}[\boldsymbol{\eta}_{i}]$ and $\boldsymbol{\Lambda}_{0j} = \operatorname{diag}\left(\frac{\lambda_{1}^{4}}{\tau_{j1}^{2}},...,\frac{\lambda_{K}^{4}}{\tau_{jK}^{2}}\right)$. 

Another challenge is that the precision matrix $\boldsymbol{\Lambda}_{j}$ should be constrained within the space of $K \times K$ positive definite matrix and the traditional gradient descent may violate this constraint. In addition, it is well-known that the gradient descent on the Euclidean space does not adequately capture the geometry of the probability distribution\citep{amari1998natural}. A small Euclidean distance between $\varphi$ and $\varphi^{\prime}$ is not equivalent to a small Kullback-Leibler divergence between $q_{j}^{\varphi}(\boldsymbol{B}_{j\cdot})$ and $q_{j}^{\varphi^{\prime}}(\boldsymbol{B}_{j\cdot})$, and thus \cite{hoffman2013stochastic} suggested using the natural gradient for updating the variational parameters. The natural gradient descent can be viewed as a second-order optimization method where the Fisher information of the variational posterior takes the role of the Hessian matrix \citep{martens2020new}, resulting in faster convergence. 

Unfortunately, the Fisher information matrix of the variational parameter $(\boldsymbol{\mu}_{j},\boldsymbol{\Lambda}_{j},\nu_{j})$ from $q_{j}(\boldsymbol{B}_{j\cdot})$ has no closed-form expression. However, if we fix $\nu_{j}>2$ and using the fact that the Multivariate t-distribution has a Gaussian-Gamma mixture representation shown in (\ref{eq:Gaussian_Gamma}), then the Fisher information matrix of $(\boldsymbol{\mu}_{j},\boldsymbol{\Lambda}_{j})$ has a closed-form expression in the joint distribution $q_{j}(\boldsymbol{B}_{j\cdot},\boldsymbol{u}_{j})$, such that 
$$
I_F(\boldsymbol{\mu}_{j},\boldsymbol{\Lambda}_{j})=\left[\begin{array}{cc} \boldsymbol{\Lambda}_{j} & 0 \\ 0 &  \frac{1}{2}\boldsymbol{\Lambda}_{j}^{-1} \otimes \boldsymbol{\Lambda}_{j}^{-1} \end{array}\right],
$$
where $\otimes$ is the Kronecker product. That's the reason why we update $\boldsymbol{\nu}$ separately. The next theorem provides a data augmentation perspective that justifies the above argument.
\begin{theorem}\label{thm:data_augmentation_perspective}
With the mean filed assumption $q(\boldsymbol{\boldsymbol{B}},\boldsymbol{\Omega},\boldsymbol{\eta}) = \prod_{j=1}^{p}q_{j}(\boldsymbol{B}_{j\cdot})\prod_{j=1}^{p}q_{j}(\sigma_{j}^{2}) \prod_{i=1}^{n} q_{i}(\boldsymbol{\eta}_{i})$ and $q_{j}(\boldsymbol{B}_{j\cdot}):=t_{\nu_{j}}(\boldsymbol{\mu}_{j}, \boldsymbol{\Sigma}_{j})$, then for any fixed $\nu_{j} >2$ for $j=1,...,p$, minimizing
$$\operatorname{KL}\left(\prod_{j=1}^{p}q_{j}(\boldsymbol{B}_{j\cdot})\prod_{j=1}^{p}q_{j}(\sigma_{j}^{2}) \prod_{i=1}^{n} q_{i}(\boldsymbol{\eta}_{i}) \Big{\|} \pi(\boldsymbol{B},\boldsymbol{\Omega}, \boldsymbol{\eta} \mid \boldsymbol{X})\right)$$
is equivalent to minimizing 
$$\operatorname{KL}\left(\prod_{j=1}^{p}q_{j}(\boldsymbol{B}_{j\cdot} \mid u_{j})\prod_{j=1}^{p}q(u_{j})\prod_{j=1}^{p}q_{j}(\sigma_{j}^{2}) \prod_{i=1}^{n} q_{i}(\boldsymbol{\eta}_{i}) \Big{\|} \prod_{j=1}^{p}q(u_{j}) \pi(\boldsymbol{B},\boldsymbol{\Omega}, \boldsymbol{\eta} \mid \boldsymbol{X})\right).$$
In other words, the difference of their corresponding ELBO is only a constant. 
\end{theorem}
Then we have the natural gradient 
\begin{equation}\label{eq:NG_gradient}
\begin{aligned}
I_F^{-1}(\boldsymbol{\mu}_{j})\nabla_{\boldsymbol{\mu}_{j}}\tilde{\mathcal{L}}(q_{j}(\boldsymbol{B}_{j\cdot}))  & = \boldsymbol{\Lambda}_{j}^{-1}\left[c_{j}\boldsymbol{X}_{\cdot j}^{\top}-\boldsymbol{\mu}_{j} \left[c_{j}(\boldsymbol{w}\boldsymbol{w}^{\top}+ n\Phi)+\mathbb{E}_{\pi(\boldsymbol{\tau}_{j}^{2},\boldsymbol{\lambda} \mid  \boldsymbol{B}_{j\cdot}^{*})}\left[\boldsymbol{\Lambda}_{0j}\right]\right]\right],\\
I_F^{-1}(\boldsymbol{\Lambda}_{j})\nabla_{\boldsymbol{\Lambda}_{j}}\tilde{\mathcal{L}}(q_{j}(\boldsymbol{B}_{j\cdot})) & =  \frac{\nu_{j}}{\nu_{j}-2}\left[c_{j}(\boldsymbol{w}\boldsymbol{w}^{\top}+n \Phi)+\mathbb{E}_{\pi(\boldsymbol{\tau}_{j}^{2},\boldsymbol{\lambda} \mid \boldsymbol{B}_{j\cdot}^{\star})}\left[\boldsymbol{\Lambda}_{0j} \right]\right] - \boldsymbol{\Lambda}_{j}.
\end{aligned}
\end{equation}
At each iteration $t$, we update the variational parameters by natural gradient descent
$$
\begin{aligned}
\boldsymbol{\mu}_{j,t} & = \boldsymbol{\mu}_{j,t-1} + \rho_{t}I_F^{-1}(\boldsymbol{\mu}_{j})\nabla_{\boldsymbol{\mu}_{j}}\tilde{\mathcal{L}}(q_{j}(\boldsymbol{B}_{j\cdot})) \\
\boldsymbol{\Lambda}_{j,t} & = (1 -\rho_{t})\boldsymbol{\Lambda}_{j,t-1} + \rho_{t} \frac{\nu_{j}}{\nu_{j}-2}\left[c_{j}(\boldsymbol{w}\boldsymbol{w}^{\top}+n \Phi)+\mathbb{E}_{\pi(\boldsymbol{\tau}_{j}^{2},\boldsymbol{\lambda} \mid \boldsymbol{B}_{j\cdot}^{\star})}\left[\boldsymbol{\Lambda}_{0j} \right]\right],
\end{aligned}
$$
where $\rho_{t}$ is a step size parameter such that $\sum_{t=1}^{+\infty} \rho_t \rightarrow \infty$ and $\sum_{t=1}^{+\infty} \rho_t^{2} < \infty$ \citep{robbins1951stochastic}. We see that with natural gradient descent, the positive definiteness of $\boldsymbol{\Lambda}_{j}$ can be guaranteed as long as its initialization is positive definite.

\subsubsection{Learning \texorpdfstring{$\nu_{j}$}{nu\_j}}
Since we require $\nu_{j}>2$ to guarantee that the mean and covariance of the multivariate t-distribution $q_{j}(\boldsymbol{B}_{j})$ exist, for ease of optimization, we set $ o_{j} =\nu_{j} - 2$ with $o_{j}>0$. Then the gradient of ELBO with respect to $o_{j}$ can be written as 
\begin{equation}\label{eq:gradient_v}
\begin{aligned}
\nabla_{o_{j}}\mathcal{L}_{o_{j}} & = \frac{c_{j}}{o_{j}^{2}}\left[\operatorname{Trc}[\boldsymbol{\Lambda}_{j}^{-1}(\boldsymbol{w}\boldsymbol{w}^{\top}+n\Phi)]\right] + \nabla_{o_{j}} \mathbb{E}_{q(\boldsymbol{B} \mid \boldsymbol{u})q(\boldsymbol{u})}[\log \pi(\boldsymbol{B})] \\
& + \frac{K}{2 (o_{j}+2)}+\frac{o_{j}+2+K}{4}\left[\psi^{\prime}\left(\frac{o_{j}+2+K}{2}\right)-\psi^{\prime}\left(\frac{o_{j}+2}{2}\right)\right],
\end{aligned}
\end{equation}
where $\psi^{\prime}(\cdot)$ is the trigamma function. Since standard gradient descent is inherently "Euclidean", whose update rule has no mechanism to respect constraints $o_{j}>0$, we use mirror descent \citep{beck2003mirror,raskutti2015information}, which reinterprets the update in the "dual space" with no constraint and uses the mirror map to pull the point back into the primal space. For parameter space with positive constraint, we consider a mirror map that $\phi(o_{j}) = o_{j} \log(o_{j})$, which is the generalized Shannon entropy. Its gradient function and the inverse of gradient function are $\theta_{j}=\nabla_{o_{j}}\phi(o_{j}) = 1 + \log(o_{j})$ and $o_{j}= \phi^{\star}(\theta_{j}) = \exp(\theta_{j}-1)$, respectively. At each iteration, we will
\begin{itemize}
\tightlist
    \item Compute gradient $\nabla_{o_{j}}\mathcal{L}_{o_{jt}}$.
    \item Update in the dual space : $\theta_{j,t+1}=\nabla_{o_{j}} \Phi\left(o_{j,t}\right)+\rho_{t} \nabla_{o_{j}}\mathcal{L}_{o_{jt}}$.
    \item Map back from the dual space to the primal space: $o_{j,t+1} = \phi^{\star}(\theta_{j,t+1}) = \exp(\theta_{j,t+1}-1)$.
\end{itemize}
The above steps can be combined into a very clean and simple form, which gives us the exponentiated gradient update rule:
\begin{equation}\label{eq:mirror_descent}
\boldsymbol{o}_{t+1}=\boldsymbol{o}_t \odot \exp \left(\rho_{t} \nabla_{\boldsymbol{o}}\mathcal{L}_{\boldsymbol{o}_{t}}\right).
\end{equation}
We see that the second term in (\ref{eq:gradient_v}) has no closed form. We can use the REINFORCE trick or the Reparameterization Trick to calculate the Monte Carlo gradient. In practice, gradients estimated from the reparameterization trick often have smaller variance than the REINFORCE trick, so they are preferred as long as such reparameterization exists. Unfortunately, the parameter $o_{j}$ we want to optimize is in the Gamma distribution $q_{j}(u_{j}) \sim \operatorname{Gamma}(\frac{o_{j}}{2} + 1, \frac{o_{j}}{2} + 1 )$, which is not amenable to standard reparameterization. \cite{naesseth2017reparameterization} proposed generalized reparameterization gradients for the shape parameter in the Gamma distribution and showed their superior performance over the REINFORCE trick. Here we follow their approach with small modifications adapted to our problem. By equation (\ref{eq:Gaussian_Gamma}), we write  $h(\boldsymbol{s},\boldsymbol{o},\boldsymbol{\epsilon})=(h_{1}(s_{1},o_{1},\boldsymbol{\epsilon}_{1}),...,h_{p}(s_{p},o_{p},\boldsymbol{\epsilon}_{p}))$ with 
$$
h_{j}(s_{j},o_{j}, \boldsymbol{\epsilon}_{j}) = \boldsymbol{\mu}_{j}+ \sqrt{\frac{o_{j}+2}{2}}s_{j}^{-1/2}\boldsymbol{\Lambda}_{j}^{-1/2}\boldsymbol{\epsilon}_{j},
$$
where $\boldsymbol{\epsilon}$ is $p \times K$  $i.i.d.$ standard normal distribution and $\boldsymbol{s}=(s_{1},...,s_{p})$ with $s_{j} \sim \mathrm{Gamma}(\frac{o_{j}+2}{2},1)$.  If there exists an invertible transform $s_{j} = T_{o_{j}}(z_{j})$,  then there exists $q_{\boldsymbol{o}}(\boldsymbol{z})$, such that
$$
\mathbb{E}_{q(\boldsymbol{B} \mid \boldsymbol{u})q(\boldsymbol{u})}[\log \pi(\boldsymbol{B})] = \mathbb{E}_{q(\boldsymbol{\epsilon})q_{\boldsymbol{o}}(\boldsymbol{z})}[\log \pi(h(T_{\boldsymbol{o}}(\boldsymbol{z}),\boldsymbol{o},\boldsymbol{\epsilon})].
$$
Then we can compute its gradient with respect to $\boldsymbol{o}$
\begin{equation}
\begin{aligned}
\nabla_{\boldsymbol{o}}\mathbb{E}_{q(\boldsymbol{\epsilon})q_{\boldsymbol{o}}(\boldsymbol{z})}[\log \pi(h(T_{\boldsymbol{o}}(\boldsymbol{z}),\boldsymbol{o},\boldsymbol{\epsilon})] & = \underbrace{\mathbb{E}_{q(\boldsymbol{\epsilon})q_{\boldsymbol{o}}(\boldsymbol{z})}[\nabla_{\boldsymbol{o}}\log \pi(h(T_{\boldsymbol{o}}(\boldsymbol{z}),\boldsymbol{o},\boldsymbol{\epsilon})]}_{\text {pathwise gradient }}\\
& + \underbrace{\mathbb{E}_{q(\boldsymbol{\epsilon})q_{\boldsymbol{o}}(\boldsymbol{z})}[   \nabla_{\boldsymbol{o}}\log q_{\boldsymbol{t}}(z)\log \pi(h(T_{\boldsymbol{o}}(\boldsymbol{z}),\boldsymbol{o},\boldsymbol{\epsilon})]}_{\text{score-function gradient}}.
\end{aligned}
\end{equation}
By finding an appropriate invertible transform $T_{o_{j}}(z_{j})$, the score-function term will be very small. In this case, the generalized reparameterization gradients will be close to the reparameterization gradients. \cite{naesseth2017reparameterization} consider the invertible transform
\begin{equation}\label{eq:IT}
T_{o_{j}}(z_{j})=\left(\frac{o_{j}}{2}+\frac{2}{3}\right)\left(1+\frac{z_{j}}{\sqrt{4.5 o_{j}+6}}\right)^3.
\end{equation}
By accept-reject sampling, we can generate a random variable $z_{j}$ with density
\begin{equation}\label{eq:q_z}
q_{o_{j}}(z_{j})= \frac{1}{\Gamma\left(\frac{o_j+2}{2}\right)} T_{o_j}^{\frac{o_j+2}{2}-1}(z_{j})e^{-T_{o_j}(z_j)} \left|\frac{d T_{o_j}(z_j)}{d z_j}\right| 1_{z_{j}>-\sqrt{4.5 o_{j}+6}}.
\end{equation}
Then $T_{o_{j}}(z_{j}) \sim \mathrm{Gamma}(\frac{o_{j}+2}{2},1)$. The details of both accept-reject sampling and the derivation of the gradient are given in the supplementary material. The pseudocode for our variational inference algorithm is provided in Algorithm \ref{alg:VI}.

\begin{algorithm}[H]
\label{alg:VI}
\SetAlgoLined
    \textbf{Inputs:} $\boldsymbol{X}$, $a$, $c_{1}$, $c_{2}$\\
    
\While{$|\mathcal{L}_{t}-\mathcal{L}_{t-1}| > \epsilon$}{
        $q_{i}(\boldsymbol{\eta}_{i}) \sim \mathcal{N}_{K}( (\boldsymbol{I}_{K} + \boldsymbol{\mu}^{\top}\boldsymbol{C}\boldsymbol{\mu}+ \boldsymbol{A})^{-1}\boldsymbol{\mu}^{\top}\boldsymbol{C}\boldsymbol{x}_{i},(\boldsymbol{I}_{K} + \boldsymbol{\mu}^{\top}\boldsymbol{C}\boldsymbol{\mu}+ \boldsymbol{A})^{-1})$.\\
        ${q_{j}(\sigma_{j}^{2})  \sim \operatorname{IG}\left(\alpha_{\sigma}+\frac{n}{2}, b_{\sigma}+\frac{1}{2}\|\boldsymbol{X}_{\cdot j}^{\top} - \boldsymbol{\mu}_{j \cdot}\boldsymbol{w} \|^{2} +\frac{n}{2}\boldsymbol{\mu}_{j}\Phi \boldsymbol{\mu}_{j}^{\top} +\frac{1}{2}\operatorname{Trc}\left[\operatorname{Cov}(\boldsymbol{B}_{j \cdot})(\boldsymbol{w}\boldsymbol{w}^{\top}+n\Phi)\right] \right)}$.\\
        Updating $(\boldsymbol{\mu}_{j},\boldsymbol{\Lambda}_{j})$ via natural gradient descent:\\
\For{$t \leftarrow 1$ \KwTo $T_{1}$}{
        $$
        \begin{aligned}
        \boldsymbol{\mu}_{j,t} & = \boldsymbol{\mu}_{j,t-1} + \rho_{t}I_F^{-1}(\boldsymbol{\mu}_{j})\nabla_{\boldsymbol{\mu}_{j}}\tilde{\mathcal{L}}(q_{j}(\boldsymbol{B}_{j})) \\
        \boldsymbol{\Lambda}_{j,t} & = (1 -\rho_{t})\boldsymbol{\Lambda}_{j,t-1} + \rho_{t} \frac{\nu_{j}}{\nu_{j}-2}\left[c_{j}(\boldsymbol{w}\boldsymbol{w}^{\top}+n \Phi)+\operatorname{E}_{\pi(\boldsymbol{\tau}_{j}^{2},\boldsymbol{\lambda} \mid \boldsymbol{B}_{j}^{\star})}\left[\boldsymbol{\Lambda}_{0j} \right]\right]
        \end{aligned}
        $$}
        Set $\boldsymbol{o} = \boldsymbol{\nu}-2$ and calculate the MC gradient $\nabla_{\boldsymbol{o}}\hat{\mathcal{L}}_{\boldsymbol{o}}$.\\
           \For{$t \leftarrow 1$ \KwTo $T_{2}$}{
            $\boldsymbol{o}_{t+1}=\boldsymbol{o}_t \odot \exp \left(\rho_{t} \nabla_{\boldsymbol{o}}\hat{\mathcal{L}}_{\boldsymbol{o}}\right)$}
            Set $\boldsymbol{\nu} = \boldsymbol{o} +2$}
            
\KwResult{$\prod_{j=1}^{p}q_{j}(\boldsymbol{B}_{j \cdot})$, $\prod_{j=1}^{p}q_{j}(\sigma_{j}^{2})$ and  $\prod_{i=1}^{n} q_{i}(\boldsymbol{\eta}_{i})$ }
\caption{Variational Inference}
\end{algorithm}

\section{Numerical studies}

We demonstrate the performance of the proposed algorithms for the Bayesian factor model with $L_{1/2}$ shrinkage using two synthetic examples and two gene datasets. We compared our algorithms with other competing methods, including the Gibbs sampler(GS) and variational inference(VI) for MGP shrinkage prior \citep{bhattacharya2011sparse,hansen2025fast}, the Gibbs Sampler for CSP prior \citep{legramanti2020bayesian} and the parameter expanded likelihood expectation maximization (PXL-EM) for SSL-IBP \citep{rovckova2016fast}. Throughout the numerical studies, we set $a = 15$, $c_{1} = 2.3$ and $c_{2} = 0.7$ for our $L_{1/2}$ shrinkage prior. All the hyperparameters involved in the competitors are set as suggested by their reference. All the Gibbs samplers are run for 10,000 iterations with the first 5,000 iterations discarded as a burn-in period. In addition, we always set $K =50$ for all the methods in this section except for the variational inference for MGP \citep{hansen2025fast}. We encountered numerical instability issues with this algorithm when setting $K>30$. Thus, we set $K=30$ for this algorithm. The increasing shrinkage priors will allow us to estimate the effective dimension of the latent factor $\hat{K}$, which is often much smaller.   

\subsection{Simulation studies}
We consider two synthetic examples that have been studied by \cite{rovckova2016fast} and \cite{hansen2025fast} respectively. These two examples incorporate sparsity and specific structural patterns in the loading matrix, leading to sparsity structures in the non-diagonal elements of the covariance matrix. In these two examples, we will repeat the experiments for $50$ times and report the mean and the standard deviation of the following metrics: 
\begin{itemize}
\tightlist
    \item The Frobenius norm between $\boldsymbol{\Sigma}_{0}$ and $\hat{\boldsymbol{\Sigma}}$.
    \item The false discovery rate (FDR) and the false negative rate (FNR) of the nonzero elements in the covariance matrix.
    \item The estimated effective factor cardinality. 
\end{itemize}
It should point out that, except for the SSL-IBP, we can not obtain the sparsity of the factor loading matrix $\boldsymbol{B}$ directly from the other methods as the continuous shrinkage priors assign zero probability to the point zero. Thus, some post selection inference procedures are required. To estimate the effective latent dimension, we first estimate the sparsity of the factor loading matrix by checking for zero in marginal credible intervals \citep{van2017uncertainty}. If $0$ is not in the marginal credible interval, we declare it a signal, and otherwise we declare it as zero. If all the elements in the column vector $\boldsymbol{B}_{\cdot k}$ are identified as zero, then the latent dimension $k$ is not counted as effective. We will use $95\%$ credible intervals in our numerical studies. To estimate the sparsity structure of the covariance, we use the simple thresholding rule: the element in the covariance matrix will be identified as zero if the absolute value of its posterior mean is smaller than $10^{-4}$.

\subsubsection{Synthetic example 1}

\begin{figure}[htbp]
    \centering
    \begin{subfigure}[b]{0.325\textwidth}
        \centering
        \includegraphics[width=\linewidth]{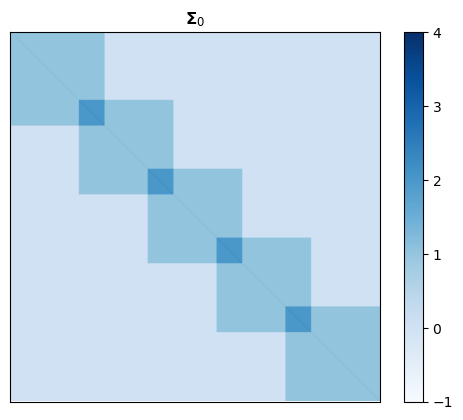}
        \caption{True}
    \end{subfigure}
    \hfill 
    \begin{subfigure}[b]{0.325\textwidth}
        \centering
        \includegraphics[width=\linewidth]{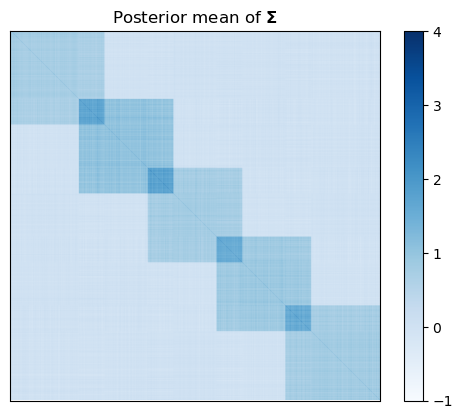}
        \caption{Gibbs Sampler}
    \end{subfigure}
    \hfill 
    \begin{subfigure}[b]{0.325\textwidth}
        \centering
        \includegraphics[width=\linewidth]{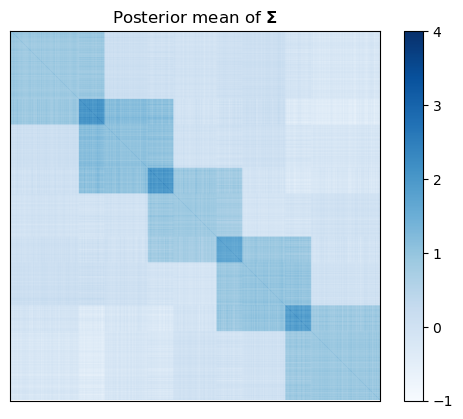}
        \caption{Variational Inference}
    \end{subfigure}
    \caption{(a) The true structure of the covariance matrix, (b) the posterior mean of the covariance matrix from Gibbs sampler with $L_{1/2}$ shrinkage, (c) the posterior mean of the covariance matrix from variational inference with $L_{1/2}$ shrinkage.}
    \label{fig:cov}
\end{figure}

For the first synthetic example, the data is generated by drawing $n = 100$ independent samples from the multivariate normal distribution $\mathcal{N}_{p}(\boldsymbol{0},\boldsymbol{B}_{0}\boldsymbol{B}_{0}^{\top}+\boldsymbol{I}_{p})$ with $p = 1956$. Assuming $K_{0}=5$ latent factors, the $p \times K_{0}$ factor loading matrix $\boldsymbol{B}_{0}$ follows a block-diagonal structure with overlapping nonzero elements allocation. Specifically, each factor is associated with $500$ nonzero responses, such that $\sum_{j=1}^{p} 1_{\left\{B_{jk} \neq 0\right\}} = 500$, while the overlap between successive factors is $\sum_{j=1}^{p} 1_{\left\{B_{jk}B_{j,k+1} \neq 0\right\}} = 136$. The value of all the nonzero elements in the factor loading matrix are set to be 1. The block-diagonal structure of $\boldsymbol{B}_{0}$ leads to a true covariance matrix that is also block-diagonal, as seen in Figure \ref{fig:cov}(a). 

Figure \ref{fig:cov}(b) and Figure \ref{fig:cov}(c) show that both our two inference approaches provide accurate estimation for the covariance matrix. The GS performed better than the VI as Figure \ref{fig:cov}(c) is slightly blurred in the true sparsity area, while Figure \ref{fig:cov}(b) is quite clean. Table \ref{tab:comparison_s1} confirms this observation as the Frobenius norm from VI is larger than the GS. Comparing with the competitors, both inference approaches for $L_{1/2}$ shrinkage consistently outperform the MGP and CSP but are beaten by the SSL-IBP from \cite{rovckova2016fast}, which achieved smaller FDR and Frobenius norm. 

\begin{table}[htbp]
  \centering
  \begin{tabular}{lccccc}
    \toprule
    & & \textbf{FDR} & \textbf{FNR} & \boldmath$\hat{K}$ & \textbf{Frobenius norm} \\
    \midrule
    \multirow{2}{*}{$L_{1/2}$}
      & GS  & 69.28\% (0.00)   & 0.00\% (0.00)       & 5.00 (0.00)       & 344.79 (39.11) \\
      & VI     & 69.28\% (0.00)   & 0.00\% (0.00)       & 5.00 (0.00)       & 416.12 (42.42) \\
    \midrule
    \multirow{2}{*}{MGP}
      & GS  & 69.28\% (0.00)   & 0.00\% (0.00)       & 5.00 (0.00)       & 442.42 (41.52) \\
      & VI     & 68.50\% (0.00)   & 0.01\% (0.00)  & 5.00 (0.00)       & 597.03 (31.47) \\
    \midrule
    CSP
      & GS  & 69.29\% (0.00)   & 0.00\% (0.00)       & 5.00 (0.00)       & 487.70 (53.59) \\
    \midrule
    SSL-IBP
      & PXL-EM &  \, 7.10\% (0.09)  & 0.08\% (0.00)  & 5.04 (0.20) & 248.52 (44.55) \\
    \bottomrule
  \end{tabular}
  \caption{Comparison of inference algorithms for different priors.}
   \label{tab:comparison_s1}
\end{table}

\begin{figure}[htbp]
    \centering
    \begin{subfigure}[b]{0.325\textwidth}
        \centering
        \includegraphics[width=\linewidth]{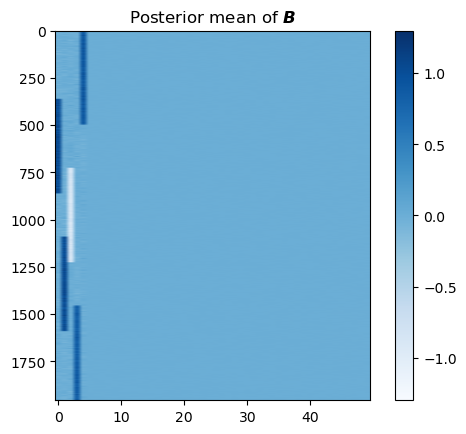}
        \caption{$L_{1/2}$}
    \end{subfigure}
    \hfill 
    \begin{subfigure}[b]{0.325\textwidth}
        \centering
        \includegraphics[width=\linewidth]{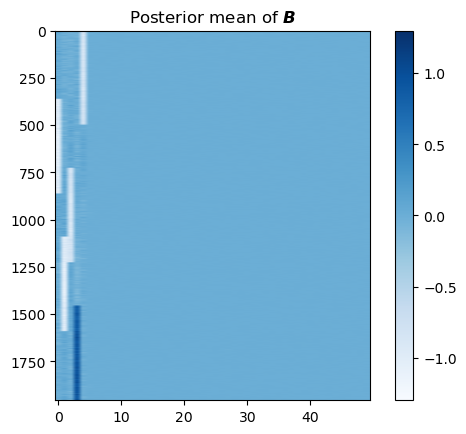}
        \caption{MGP}
    \end{subfigure}
    \hfill 
    \begin{subfigure}[b]{0.325\textwidth}
        \centering
        \includegraphics[width=\linewidth]{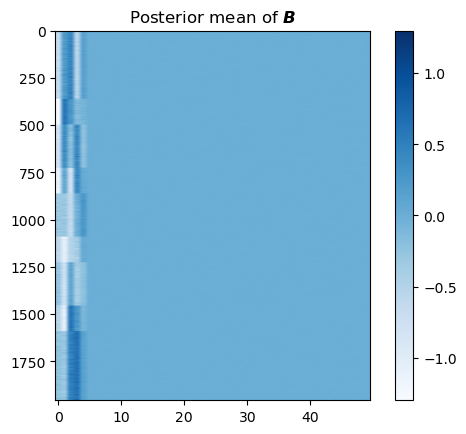}
        \caption{CSP}
    \end{subfigure}
    \caption{The heatmap of the posterior mean from the Gibbs samplers for the loading matrix $\boldsymbol{B}$ with three different shrinkage priors.}
    \label{fig:B_mean}
\end{figure}

Figure \ref{fig:B_mean} provides the heatmap of the posterior mean from the GS for the loading matrix $\boldsymbol{B}$ with three different shrinkage priors. We see that for all three methods, the value of the posterior mean are close to zero for elements in the factor loading matrix with column index larger than five. Within the first five latent dimensions, both $L_{1/2}$ shrinkage and MGP shrinkage provide a more structured sparsity estimation than CSP. But due to the identifiable issue, none of them recover the exact factor loading matrix. 

\begin{figure}[htbp]
    \centering
    \begin{subfigure}[b]{0.325\textwidth}
        \centering
        \includegraphics[width=\linewidth]{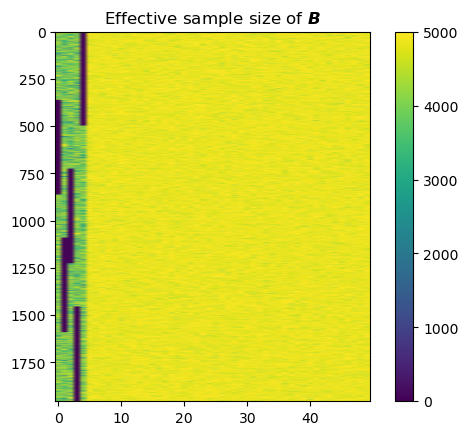}
        \caption{$L_{1/2}$}
    \end{subfigure}
    \hfill 
    \begin{subfigure}[b]{0.325\textwidth}
        \centering
        \includegraphics[width=\linewidth]{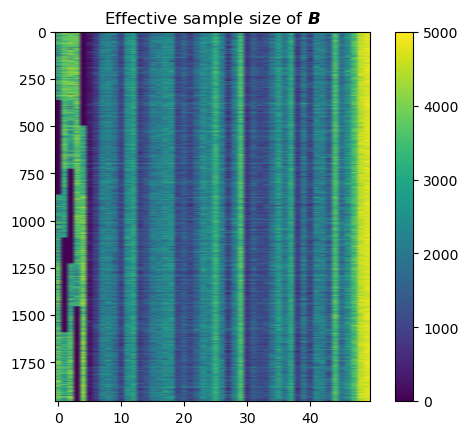}
        \caption{MGP}
    \end{subfigure}
    \hfill 
    \begin{subfigure}[b]{0.325\textwidth}
        \centering
        \includegraphics[width=\linewidth]{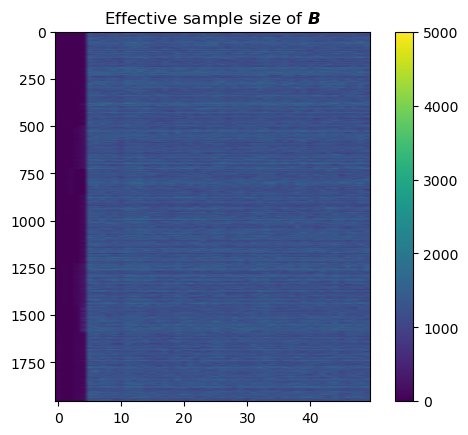}
        \caption{CSP}
    \end{subfigure}
    \caption{The heatmap of the effective sample size from the Gibbs samplers for the loading matrix $\boldsymbol{B}$ with three different shrinkage priors.}
    \label{fig:ess_B}
\end{figure}

We also compare the convergence speed of GS in this example. This is done by calculating the effective sample size (ESS) for $\boldsymbol{B}$ after the burn-in period using \textbf{arviz} package \citep{kumar2019arviz} and plotting their heatmaps in Figure \ref{fig:ess_B}. In terms of ESS, the GS for $L_{1/2}$ shrinkage has the best performance. Interestingly, the heatmap of the ESS for $\boldsymbol{B}$ are somehow consistent with the heatmap of its posterior mean, particularly for the $L_{1/2}$ shrinkage. 
The ESS are large for elements in $\boldsymbol{B}$ that belong to the non-effective latent dimensions.  For $L_{1/2}$ shrinkage, the autocorrelation of the samples in these elements is almost zero. On the other hand, the ESS is very small for elements identified as nonzero.

\subsubsection{Synthetic example 2}

For the second synthetic example, the data is generated from $\mathcal{N}_{p}(\boldsymbol{0},\boldsymbol{B}_{0}\boldsymbol{B}_{0}^{\top}+\boldsymbol{\Omega}_{0})$ with three $(n, p)$ combinations, namely $(100,1000)$, $(500,1000)$ and $(500,5000)$. With the number of latent factors $K_{0}=5$, each entry of the loading matrix $\boldsymbol{B}_{0}$ was set to zero with a probability $2/3$, or drawn from $\mathcal{U}(0,1)$ distribution with a probability of $1/3$.  The elements in the diagonal matrix $\boldsymbol{\Omega}_{0}$ were sampled independently from $\mathcal{U}(0.1,1)$. 

Table \ref{tab:comparison_s2} summarizes the numerical results from three scenarios. For Scenario 2 and Scenario 3, both inference approaches from $L_{1/2}$ shrinkage always outperformed all competitors. For Scenario 1, the $L_{1/2}$ shrinkage still outperformed the MGP shrinkage prior and CSP prior but had a larger Frobenius norm than SSL-IBP. SSL-IBP had a smaller FDR but a much larger FNR than the other methods, indicating that SSL-IBP overestimates the sparsity level of the covariance matrix. Additionally, for Scenarios 3, the SSL-IBP failed to estimate the effective latent dimension as it always had nonzero elements in each column vector of $\hat{\boldsymbol{B}}$. For all the scenarios, $L_{1/2}$ shrinkage can target the right number of the effective latent dimension exactly. 

\begin{table}[htbp]
  \centering
  \begin{tabular}{lccccc}
    \toprule
    & & \textbf{FDR} & \textbf{FNR} & \boldmath$\hat{K}$ & \textbf{Frobenius norm} \\
    \midrule
    
    \multicolumn{6}{c}{\text{Scenario 1: $n = 100, \; p = 1000$}} \\
    \cmidrule(lr){1-6}

    \multirow{2}{*}{$L_{1/2}$} 
      & GS  & 55.20\% (0.01) & 0.07\% (0.00) & 5.00 (0.00)       & 93.59 (8.17)   \\
      & VI     & 54.35\% (0.02) & 0.42\% (0.00) & 5.00 (0.00)       & 101.53 (10.98) \\
    \midrule 
    
    \multirow{2}{*}{MGP} 
      & GS  & 55.37\% (0.01) & 0.00\% (0.00) & 5.00 (0.00) & 134.26 (11.52) \\
      & VI     & 50.20\% (0.01) & 3.20\% (0.01) & 5.58 (0.50) & 123.81 (8.06) \\
    \midrule 
    
    CSP
      & GS  & 55.47\% (0.01) & 0.02\% (0.00) & 4.94 (0.24) & 129.40 (10.48) \\
    \midrule 
    
    SSL-IBP
      & PXL-EM & 16.10\% (0.04) & 31.70\% (0.02)& 6.72 (1.23) & 88.21 (9.34)   \\
      
    \midrule
    \multicolumn{6}{c}{\text{Scenario 2: $n = 500, \; p = 1000$}} \\
    \cmidrule(lr){1-6}
    
    \multirow{2}{*}{$L_{1/2}$} 
      & GS  & 55.32\% (0.01) & 0.06\% (0.00) & 5.00 (0.00)       & 38.21 (3.84)   \\
      & VI     & 54.98\% (0.01) & 0.05\% (0.00) & 5.00 (0.00)       & 52.24 (5.73) \\
    \midrule 
    
    \multirow{2}{*}{MGP} 
      & GS  & 55.49\% (0.01) & 0.05\% (0.00) & 5.10 (0.30)       & 65.70 (4.81) \\
      & VI     & 50.20\% (0.01) & 3.20\% (0.01) & 7.00 (0.00)       & 64.39 (5.83) \\
    \midrule 
    
    CSP
      & GS  & 55.31\% (0.01) & 0.03\% (0.00) & 5.00 (0.00) & 64.35 (5.11) \\
    \midrule 
    
    SSL-IBP
      & PXL-EM & 54.00\% (0.01) & 0.50\% (0.00)& 42.16 (6.28) & 93.61 (8.28)   \\

    \midrule
    \multicolumn{6}{c}{\text{Scenario 3: $n = 500, \; p = 5000$}} \\
    \cmidrule(lr){1-6}
    
    \multirow{2}{*}{$L_{1/2}$} 
      & GS  & 55.13\% (0.01) & 0.06\% (0.00) & 5.00 (0.00)       & 195.46 (20.67)   \\
      & VI     & 54.98\% (0.01) & 0.06\% (0.00) & 5.00 (0.00)       & 262.62 (17.94) \\
    \midrule 
    
    \multirow{2}{*}{MGP} 
      & GS  & 55.25\% (0.01) & 0.05\% (0.00) & 5.06 (0.24)       & 374.56 (17.56) \\
      & VI     & 45.60\% (0.01) & 3.20\% (0.00) & 7.26 (0.44) & 348.97 (23.49) \\
    \midrule 
    
    CSP
      & GS  & 55.49\% (0.01) & 0.03\% (0.00) & 5.00 (0.00) & 373.67 (19.11) \\
    \midrule 
    
    SSL-IBP
      & PXL-EM & 55.40\% (0.01) & 0.01\% (0.00)& 50.00 (0.00) & 310.27 (39.80)   \\
    \bottomrule
  \end{tabular}
   \caption{Comparison of inference algorithms for different priors}
   \label{tab:comparison_s2}
\end{table}

Finally, we made a runtime comparison of different Bayesian approaches across varying data dimensions and hardware configurations in Table \ref{tab:runtime_comparison}. Apart from our methods, the Pytorch implementations for the GS have also been provided for both MGP and CSP, which allowing them to be compared both in CPU and GPU. For the VI for MGP and PXL-EM for SSL-IBP, we only ran the R code they provided on CPU.

On the Macbook Air M5, the GS for MGP had the least computation cost; it was significantly faster than the GS for both $L_{1/2}$ shrinkage and CSP. While on the GPU, the GS for $L _{1/2}$ shrinkage became the fastest, CSP sat in the middle, and the GS for MGP seemed to scale less efficiently than the other two on GPU. Additionally, the only difference among the three GS are the conditional posterior of global and local shrinkage parameters, suggesting that the GS for $L_{1/2}$ shrinkage has the best parallelization strategies when sampling these conditional posteriors.

The VI for $L_{1/2}$ shrinkage is consistently much faster than the GS across all platforms and data sizes. The VI for MGP shrinkage is faster than our VI approach. However, due to numerical instability issues, their approach could only be implemented with $K = 30$, whereas for the other methods we set $K =50$. The PXL-EM for SSL-IBP, which is a posterior mode search algorithm, is extremely fast even without GPU acceleration. 

The numerical studies in two synthetic examples underscore the importance of both algorithmic efficiency (MCMC vs. VI) and hardware selection (CPU vs. GPU) in scalable computation for Bayesian factor models. The GPU acceleration provides order-of-magnitude speedups for the GS. Our VI algorithm also benefits from GPU acceleration, though less dramatically. For large scale problems, our VI algorithm offers a favorable speed-accuracy trade-off.

\begin{table}[htbp]
    \centering
    \small 
    \begin{tabular}{
        l           
        l           
        S[table-format=4.0] 
        S[table-format=4.0] 
        S[table-format=4.0] 
        S[table-format=4.0] 
    }
    \toprule
    & & \multicolumn{4}{c}{Data Dimensions $(n, p)$} \\
    \cmidrule(lr){3-6}
    & & {(100, 1956)} & {(100, 1000)} & {(500, 1000)} & {(500, 5000)} \\
    \midrule

    \multicolumn{6}{c}{\textbf{Macbook Air M5 32GB RAM}} \\
    \midrule 
    
    \multirow{2}{*}{$L_{1/2}$} 
        & GS & 633   & 217   & 1226  & 7480  \\
        & VI    & 48    & 24    & 25    & 128   \\
    \midrule
    
    \multirow{2}{*}{MGP} 
        & GS & 333   & 160   & 482   & 2310  \\
        & VI    & 10    & 8     & 21  & 119   \\
    \midrule
    
    CSP & GS & 738   & 228   & 1281  & 7300  \\
    \midrule
    
    SSL-IBP & PXL-EM & 6    & 7     & 90    & 11    \\
    \midrule
    
    \multicolumn{6}{c}{\textbf{NVIDIA V100 32GB}} \\
    \midrule
    
    \multirow{2}{*}{$L_{1/2}$} 
        & GS & 120   & 85    & 143   & 462   \\
        & VI    & 34    & 27    & 28    & 57    \\
    \midrule
    
    MGP & GS & 320   & 297   & 310   & 488   \\
    \midrule
    
    CSP & GS & 141   & 98    & 160   & 479   \\
    \midrule
    
    \multicolumn{6}{c}{\textbf{NVIDIA A100 40GB}} \\
    \midrule
    
    \multirow{2}{*}{$L_{1/2}$} 
        & GS & 50    & 41    & 66    & 210   \\
        & VI    & 19    & 19    & 18    & 28    \\
    \midrule
    
    MGP & GS & 284   & 277   & 285   & 346   \\
    \midrule
    
    CSP & GS & 65    & 56    & 82    & 262   \\
    \bottomrule
    \end{tabular}
\caption{Average computation cost in seconds across 10 simulation replications for Macbook Air M5 and 50 simulation replications for NVIDIA GPUs.} 
\label{tab:runtime_comparison}
\end{table}

\subsection{Real data application}
We now present two gene datasets to illustrate our methods. 

\subsubsection{Lung cancer dataset}
We study the gene expression dataset for lung cancer, which comprises expression levels of 5000 genes and 56 subjects,  previously analyzed by \cite{lee2010biclustering} and \cite{tan2014sparse} for the biclustering task. Its subjects encompass 20 pulmonary carcinoid subjects (carcinoid), 13 colon cancer metastasis subjects (colon), 17 normal lung subjects (normal), and 6 small cell lung subjects (small cell). Due to $p>>n$ and $n$ is very small, such problem is often very challenge. Our goal here is to identify statistically significant genes correlated with lung cancer and distinguish the four different tumor types. As before, we use the posterior mean as the estimator for all the methods except for SSL-IBP, which uses the posterior mode.

\begin{figure}[ht]
    \centering
    \begin{subfigure}[b]{0.32\textwidth}
        \centering
        \includegraphics[width=\linewidth]{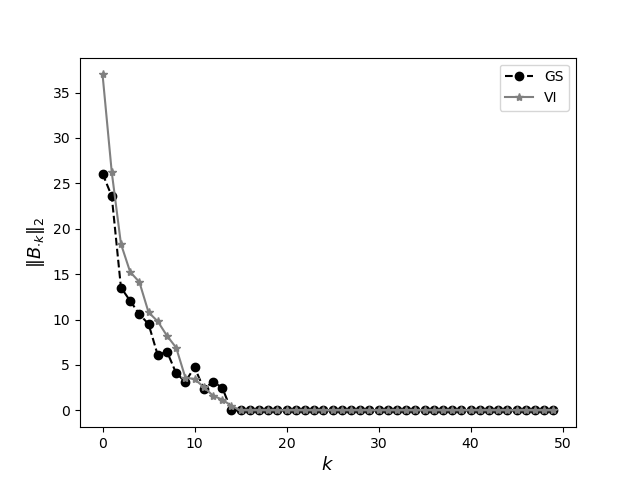}
        \caption{$L_{1/2}$ shrinkage}
    \end{subfigure}
    \hfill 
    \begin{subfigure}[b]{0.32\textwidth}
        \centering
        \includegraphics[width=\linewidth]{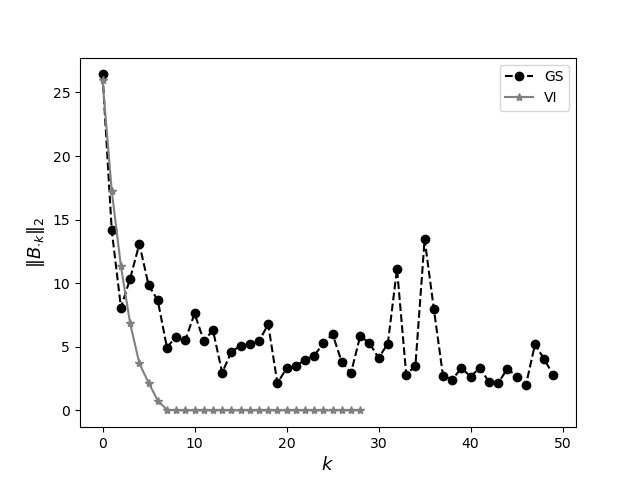}
        \caption{MGP shrinkage}
    \end{subfigure}
    \hfill 
    \begin{subfigure}[b]{0.32\textwidth}
        \centering
        \includegraphics[width=\linewidth]{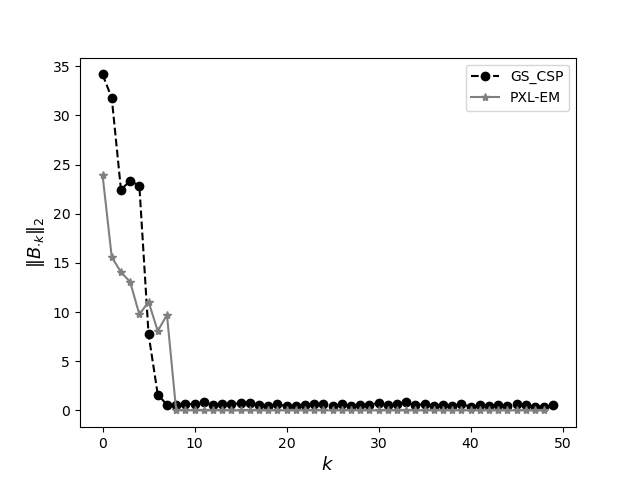}
        \caption{CSP and SSL-IBP shrinkage}
    \end{subfigure}
    \caption{$L_{2}$ norm of the posterior estimation $\hat{\boldsymbol{B}}_{\cdot k}$.}
    \label{fig:norm_B}
\end{figure}

\begin{figure}[htbp]
    \centering
    \begin{subfigure}[b]{0.49\textwidth}
        \centering
        \includegraphics[width=\linewidth]{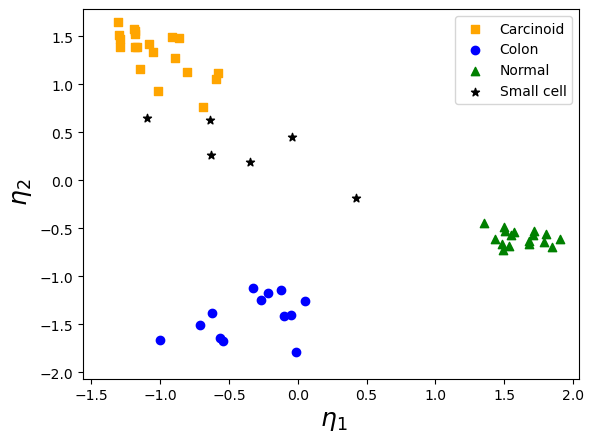}
        \caption{Gibbs Sampler}
    \end{subfigure}
    \hfill
    \begin{subfigure}[b]{0.49\textwidth}
        \centering
        \includegraphics[width=\linewidth]{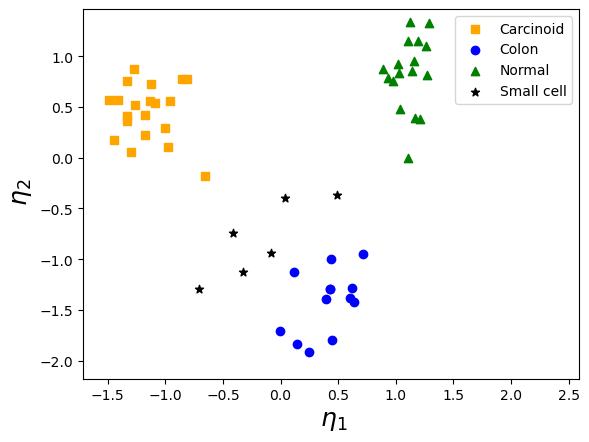}
        \caption{Variational Inference}
    \end{subfigure}
    \caption{Scatter plot for the posterior mean of $(\eta_{1}, \eta_{2})$ from $L_{1/2}$ shrinkage}
    \label{fig:score_LH}
\end{figure}

To preprocess the data for analysis, we center and scale it before running each algorithm. From Figure \ref{fig:norm_B} we observe that, although some fluctuations exist, all the methods except the GS for MGP shrinkage show a strong trend of decreasing norm of $\hat{\boldsymbol{B}}_{\cdot k}$ as the column index increases. For the $L_{1/2}$ shrinkage, the behavior of $\|\hat{\boldsymbol{B}}_{\cdot k}\|_{2}$ from GS and VI are very similar. Figure \ref{fig:score_LH} presents the posterior estimation of the latent factor scores. We see that the first two latent factors capture biologically meaningful variation that separates the four types of tumors reasonably well for our two algorithms. Similar plots from the other methods have also been given in  supplementary Appendix C. All of them separates the four types of tumors. 

To find the relevant genes correlated with lung cancer, we select the nonzero elements in the loading factor matrix $\boldsymbol{B}$ by checking for zero in the credible intervals with $95\%$ level for $L_{1/2}$ shrinkage, MGP shrinkage and CSP shrinkage. Since the PXL-EM for SSL-IBP produces the exact sparsity for the loading matrix $\boldsymbol{B}$, selecting the relevant genes can be done directly.

Table \ref{tab:nonzero_lung_cancer} lists the percentage of nonzero elements in the first six column vectors in the loading factor matrix. We observe that all methods, except GS for MGP, generally show a trend where the percentage of nonzero loadings decreases as $k$ increases from $1$ to $6$. This indicates that the early factors capture the main signal, while later factors are increasingly shrunk. In addition, the GS produced more sparsity than VI for $L_{1/2}$ shrinkage. This could be two reasons:
\begin{itemize}
\tightlist
    \item When $p>>n$ and n is small, the posterior is complex and multi-modal(zero mode and nonzero mode). VI tries to approximate it with a uni-modal distribution (i.e. multivariate-t), the optimization often faces a trade-off. To avoid the penalty for missing the probability region around nonzero mode, it may shift its posterior mean slightly away from zero. From Figure \ref{fig:score_LH}(a), we see that VI provided the larger $L_{2}$ norm of $\|\hat{\boldsymbol{B}}_{\cdot k}\|_{2}$ than GS.
    \item The mean-field assumption may lead to a narrow credible intervals as it often underestimates the uncertainty.
\end{itemize}

Thus, for problems with small sample size, we recommend using GS for $L_{1/2}$ shrinkage.

\begin{table}
\centering
\begin{tabular}{c cc cc c c}
\toprule
\multirow{2}{*}{$k$} & \multicolumn{2}{c}{$L_{1/2}$} & \multicolumn{2}{c}{MGP } & \multicolumn{1}{c}{CSP} & \multicolumn{1}{c}{SSL-IBP} \\
\cmidrule(lr){2-3} \cmidrule(lr){4-5} \cmidrule(lr){6-6} \cmidrule(lr){7-7}
& GS & VI & GS & VI & GS & PXL-EM \\
\midrule
1 & 43.14\% & 80.36\% & 17.78\% & 68.78\% & 69.88\% & 51.52\% \\
2 & 35.40\% & 71.24\% & 19.60\% & 46.98\% & 61.06\% & 34.60\% \\
3 & 22.00\% & 52.12\% & 1.58\%  & 29.06\% & 34.88\% & 32.98\% \\
4 & 18.44\% & 42.24\% & 5.02\%  & 21.60\% & 37.12\% & 27.62\% \\
5 & 20.88\% & 42.76\% & 15.28\% & 16.28\% & 33.26\% & 20.38\% \\
6 & 17.56\% & 35.62\% & 10.02\% & 10.66\% & 6.54\%  & 33.28\% \\
\bottomrule
\end{tabular}
\caption{Percentage of nonzero elements in the first six column vectors for $\boldsymbol{B}$.}
\label{tab:nonzero_lung_cancer}
\end{table}

\subsubsection{Peripheral blood mononuclear cells (PBMC) dataset}

We analyze a single cell RNA-sequencing (scRNA-seq) dataset generated on 2.7k human peripheral blood mononuclear (PBMC) cells, available in the Python package \textbf{scanpy} \citep{wolf2018scanpy}. This data consists of $n = 2,700$ single cells that were sequenced on the Illumina NextSeq 500 with $p = 32,738$ genes, which gives us a count matrix. 

The values in this matrix represent the number of molecules for each gene (i.e. $p$) that are detected in each cell (i.e. $n$). We followed the standard pre-processing data approach from \cite{stuart2019comprehensive, hao2021integrated}, widely used in scRNA-seq analysis. Its steps include the selection and filtration of cells based on quality control criteria, log-normalization and scaling, and the detection of highly variable features.
Specifically, for quality control, we filtered out the cells based on three criteria: 
\begin{itemize}
\tightlist
    \item Cells that have unique feature counts over $2,500$ or less than $200$.
    \item Genes that are detected in fewer than $3$ cells.
    \item Cells with the percentage of counts in mitochondrial genes less than $5\%$. 
\end{itemize}
These filters reduced the size of the count matrix to $2638 \times 13714$. Then,  we normalized the feature expression measurements for each cell by the total expression of each feature, multiplied this by a scale factor ($10,000$ as a default setting), and log-transformed the result. Next, we found genes that are highly expressed in some cells and lowly expressed in others.  Focusing on these genes in downstream analysis helps highlight biological signals in single-cell datasets \citep{brennecke2013accounting}. This further reduced the size of the matrix to $\boldsymbol{2638 \times 2000}$. Finally, we standardized the data. The Bayesian factor model was used to reduce the data dimension, helping us cluster the cells and identify gene markers. Results from PXL-EM for SSL-IBP are not shown for this dataset as we failed to make it work.

\begin{figure}[htbp]
    \centering
    \begin{subfigure}[b]{0.48\textwidth}
        \centering
        \includegraphics[width=\linewidth]{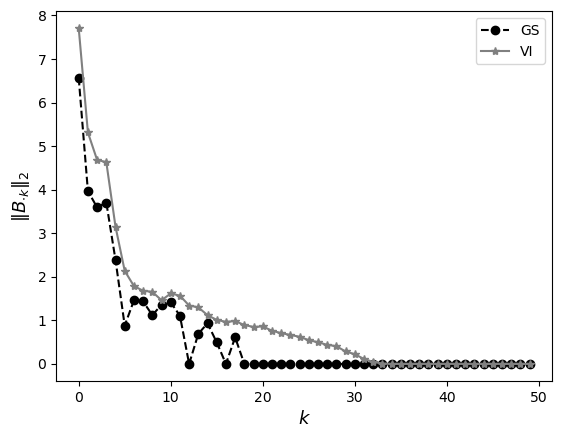}
        \caption{$L_{1/2}$ shrinkage}
    \end{subfigure}
    \hfill 
    \begin{subfigure}[b]{0.49\textwidth}
        \centering
        \includegraphics[width=\linewidth]{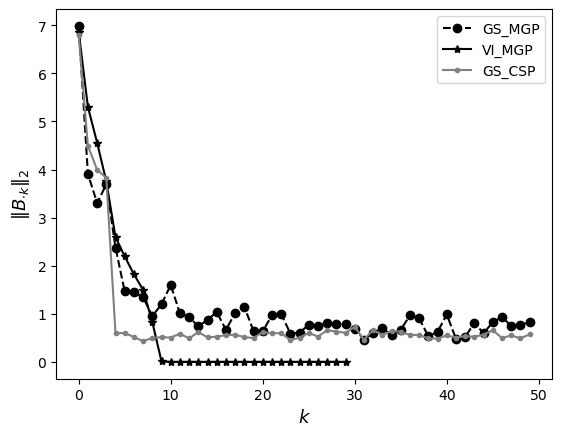}
        \caption{MGP and CSP shrinkage}
    \end{subfigure}
    \caption{The $L_{2}$ norm of the column of the posterior mean of the loading factor matrix $\boldsymbol{B}$.}
    \label{fig:norm_B_2}
\end{figure}

Figure \ref{fig:norm_B_2} shows that both the GS and VI for $L_{1/2}$ shrinkage can shrink the column of the factor loading matrix variable toward zero as its index increases, which is a desirable result in terms of dimension reduction. It allows us to use the low-dimensional latent factor scores to cluster the cell. While for its competitors, only the VI for MGP shrinkage achieves this target. The other methods only shrink the column of the factor loading matrix to a small number but not close to zero. 

With the latent factor scores, we cluster the cell  with the workflow from \cite{stuart2019comprehensive}:
\begin{itemize}
\tightlist
    \item Using posterior mean of the effective latent factor scores to compute the K-nearest neighbor (KNN) graph, with edges in the graph drawn between cells with similar score of latent factors.
    \item Clustering the neighborhood graph with the Leiden graph-clustering method \citep{traag2019louvain}.
    \item Embedding the graph in two dimensions with the UMAP algorithm \citep{mcinnes2018umap} for visualization.
\end{itemize}

\begin{figure}[htbp]
    \centering
    \begin{subfigure}[b]{0.50\textwidth}
        \centering
        \includegraphics[width=\linewidth]{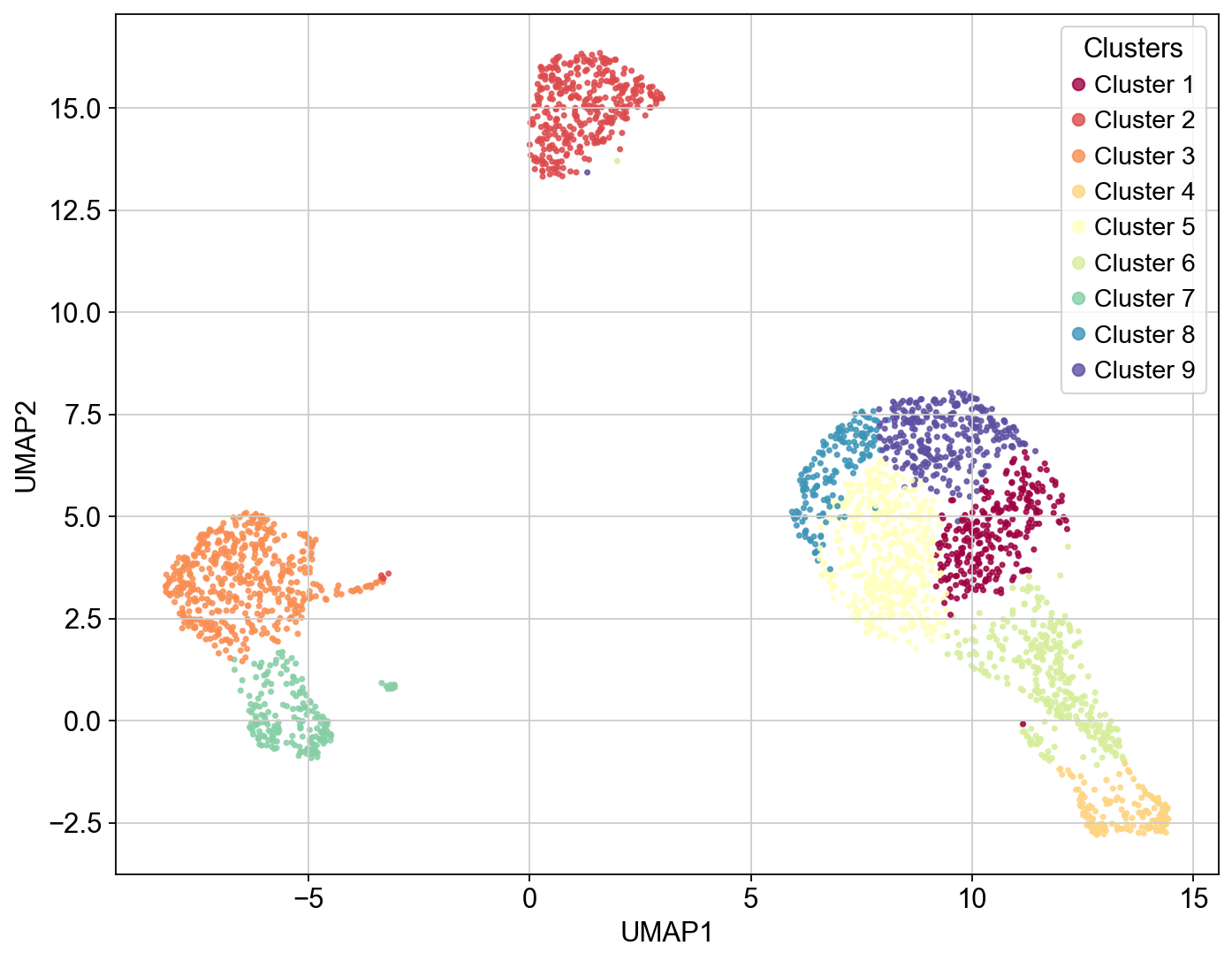}
        \caption{Gibbs Sampler}
    \end{subfigure}
    \hfill 
    \begin{subfigure}[b]{0.49\textwidth}
        \centering
        \includegraphics[width=\linewidth]{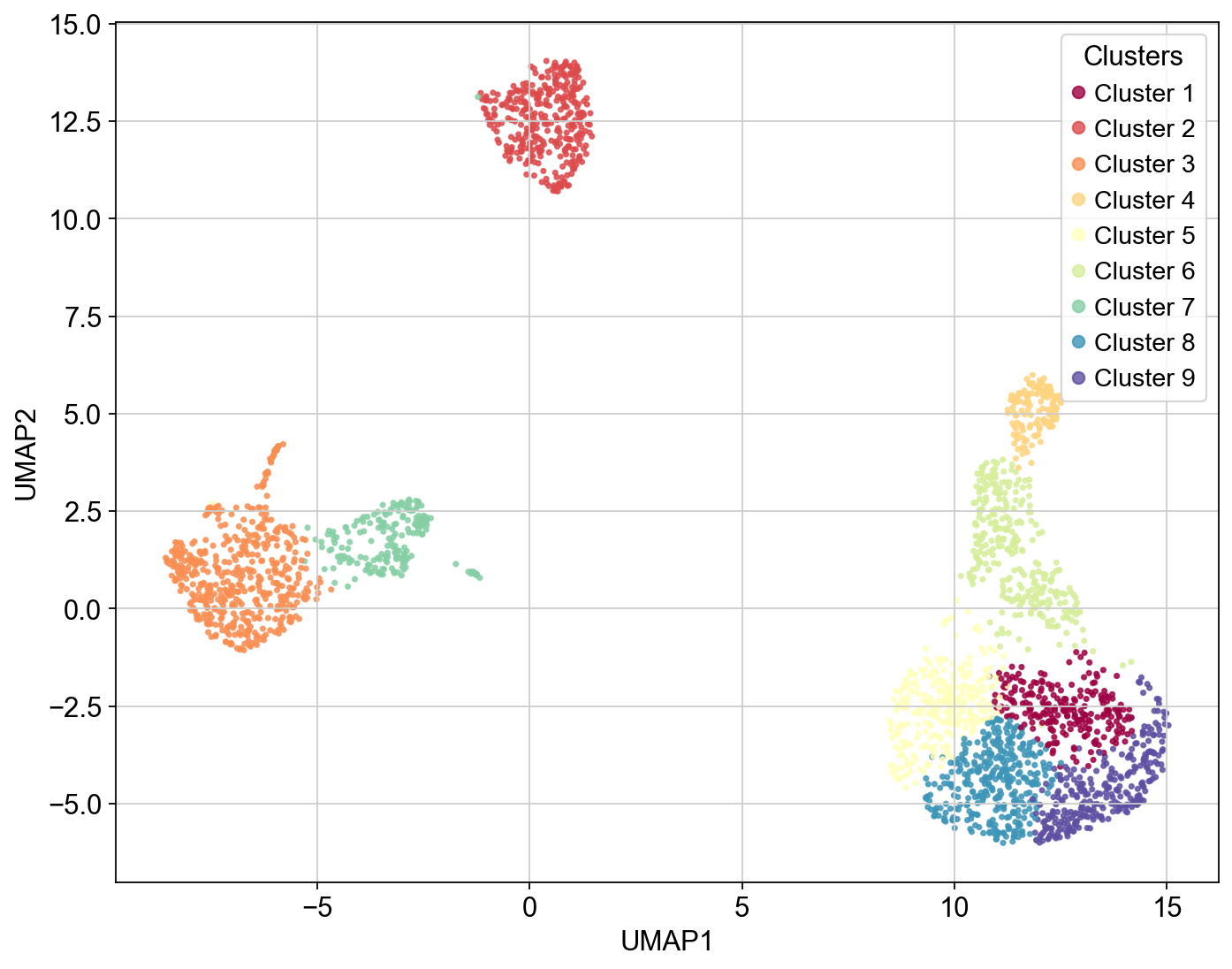}
        \caption{Variational Inference}
    \end{subfigure}
    \caption{Cell clusters from $L_{1/2}$ shrinkage visualized by projecting to two dimensions with the UMAP algorithm.}
    \label{fig:cell_cluster}
\end{figure}

From Figure \ref{fig:cell_cluster}, we see that the posterior mean of latent factor scores from the two methods for $L_{1/2}$ shrinkage yields 9 clusters, and the shape of the clusters visualized by their UMAP plots agrees quite well.

Finally, we found marker genes within each cluster, which can identify specific cell types. As suggested by \cite{ntranos2019discriminative}, the Mann–Whitney U test \citep{mann1947test} can be used to rank the marker genes in each cluster. More specifically, the test will compare the data in the cluster with the data from the rest. The rank of gene significance in the test can be obtained. Table \ref{tab:marker_LH} lists the top three marker genes in each cluster and their inferred cell type obtained by $L_{1/2}$ shrinkage. We see that the two methods give almost the same top three marker genes in each cluster. The small difference exists in cluster $1$ and cluster $5$. From Figure \ref{fig:cell_cluster} (a) and (b), we see that these two clusters are neighbors. Additionally, Table \ref{tab:marker_LH} shows that LTB and IL32 are the top three marker genes in cluster $1$ and cluster $5$. Thus, we can combine these two clusters together with inferred cell type CD4 T cells \citep{hao2021integrated}. We also see that the cluster $8$ and cluster $9$ are neighbor and their top three marker genes are ribosomal protein rich, which are constitutively expressed in all the cell types. So we will not specify their cell type. Again, these two clusters can be combined together. With these refinements, we will have 7 clusters in total. UMAP plots and the marker genes for the results from MGP shrinkage and CSP shrinkage will be given in the supplementary Appendix C. Both the results from GS and VI for MGP shrinkage prior are consistent with us. The result from GS for CSP prior only has small difference with us.

\begin{table}[htbp]
\centering
\resizebox{0.95\textwidth}{!}{%
\renewcommand{\arraystretch}{1.3} 
\label{tab:method_comparison}
\begin{tabular}{c cc cc}
\toprule
\multirow{2}{*}{\textbf{Cluster}} & \multicolumn{2}{c}{\textbf{Gibbs Sampler}} & \multicolumn{2}{c}{\textbf{Variational Inference}} \\
\cmidrule(lr){2-3} \cmidrule(lr){4-5}
& \textbf{Top Markers} & \textbf{Inferred Cell Type} & \textbf{Top Markers} & \textbf{Inferred Cell Type} \\
\midrule
1 & LTB, IL7R, LDHB & Memory T cells & LTB, IL7R, IL32 & Activated T cells \\
2 & CD74, HLA-DRA, CD79A & B cells & CD74, CD79A, HLA-DRA & B cells \\
3 & LYZ, S100A9, S100A8 & Monocytes  & LYZ, S100A9, S100A8 & Monocytes  \\
4 & GZMB, NKG7, GNLY & NK cells & GZMB, NKG7, PRF1 & NK cells  \\
5 & IL32, LDHB, LTB & Activated T cells & IL32, CD3D, LTB & Activated / Memory T cells \\
6 & CCL5, NKG7, CST7 & CD8 T cells & NKG7, CCL5, CST7 & CD8 T cells \\
7 & COTL1, LST1, FCER1G & Dendritic  & COTL1, LST1, FCER1G & Dendritic \\
8 & MALAT1, RPL32, RPS27 & Not-specific  & RPS27, RPL32, RPS12 & Not-specific  \\
9 & RPS6, RPS3A, RPS12 & Not-specific & RPS12, RPS6, RPS3A & Not-specific \\
\bottomrule
\end{tabular}%
}
\caption{Top three marker genes in each cluster and the inferred cell type obtained by $L_{1/2}$ shrinkage.}
\label{tab:marker_LH}
\end{table}

\section{Conclusion and discussion}
We have presented a new increasing shrinkage prior for discovering the number of latent factors and studied its properties. For posterior inference, we proposed the Gibbs sampler as an exact approach and variational inference as an approximation approach. We demonstrated that both our algorithms have comparable or even better performance than other algorithms discussed in the paper, showcasing their superiority.

While our VI algorithm is faster than the Gibbs sampler and scalable to large problems, it has two limitations. First, the convergence speed can be sensitive to the learning rate. Second, we can only guarantee the convergence of the ELBO, and by using a decreasing learning rate, we can also guarantee the convergence of the variational parameters. However, it is difficult to show that the convergence point is a local optimum. This is common to variational inference algorithms. Future studies should focus on understanding its convergence properties and trying to generalize our results to other models.

Lastly, an interesting research direction involves applying our increasing shrinkage prior to infinite-width neural networks. Similar to the infinite factor model, our prior would help select the width. In this context, current inference approaches cannot be applied directly due to the lack of conjugacy between the prior and likelihood. A new inference algorithm should be developed.

\section*{Disclosure statement}\label{disclosure-statement}

No potential conflict of interest was reported by the author(s).

\section{Data Availability Statement}

De-identified data have been made available at the following URL: 
\begin{itemize}
\tightlist
    \item Lung Cancer dataset: \url{https://doi.org/10.5281/zenodo.19255485}
    \item PBMC dataset: \url{https://doi.org/10.5281/zenodo.19255485}
\end{itemize}

\phantomsection\label{supplementary-material}
\bigskip

\begin{center}

{\large\bf SUPPLEMENTARY MATERIAL}

\end{center}

\begin{description}
\item[Supplementary Material:]
This supporting
information contains:
\begin{itemize}
\tightlist
    \item Appendix A. Proof of 2.1 - 2.4.
    \item Appendix B. Proof of 4.1 and 4.2.
    \item Appendix C. Some extra results in the numerical studies.
\end{itemize}
\item[Supporting code:]
The Python code is available at  \url{https://github.com/kexiongwen/Bayes_Factor_Model}
\end{description}

\bibliography{bibliography.bib}

\end{document}